\documentclass{aa}  
\usepackage{graphicx}
\usepackage{txfonts}
\usepackage{lipsum}
\usepackage{subcaption}         
\usepackage{lscape}             
\usepackage{placeins}           
\usepackage{threeparttable, tablefootnote}
\usepackage{hyperref}
\hypersetup{
    colorlinks=true,
    linkcolor=black,
    urlcolor=black,
    citecolor=blue
    }
\begin{document}
   \title{Infrared photometry and Calcium triplet spectroscopy of the most metal-poor  in-situ globular cluster VVV-CL001 \thanks{Based on observations gathered at the European Southern Observatory (programs ID 089.D-0392 and ID 091.D-0535A), at Las Campanas Observatory with the 6.5m Magellan Telescopes (program ID CN2012A-034), and with the ESO-VISTA telescope (program ID 172.B-2002).}}

   \author{W. Haro Moya\inst{1,2}
   \and C. Moni Bidin\inst{2}
   \and M.C. Parisi\inst{3,4}
   \and D. Geisler\inst{1,5}
   \and M. Bla\~na\inst{6} 
   \and S. Villanova\inst{7}
   \and F. Mauro\inst{2}
   \and A.-N. Chen\'e\inst{8}
   \and R. E. Cohen\inst{9}
   \and S. Ram\'irez Alegr\'ia\inst{10}
   \and R. Contreras Ramos\inst{11,12}
   \and M. Zoccali\inst{11,12}
   \and C. Mu\~{n}oz\inst{1}
   \and B. Dias\inst{13}
}

   \institute{ Departamento de Astronomia, Facultad de Ciencias, Universidad de La Serena. Av. Raul Bitran 1305 , La Serena, Chile
   \and Instituto de Astronom\'ia, Universidad Cat\'olica del Norte, Av. Angamos 0610, Antofagasta, Chile
   \and Observatorio Astron\'omico, Universidad Nacional de C\'ordoba, Laprida 854, X5000BGR, C\'ordoba, Argentina
   \and Instituto de Astronom\'ia Te\'orica y Experimental (CONICET-UNC), Laprida 854, X5000BGR, C\'ordoba, Argentina
   \and Departmento de Astronom\'ia, Universidad de Concepci\'on, Casilla 160-C, Concepci\'on, Chile
   \and Vicerrector\'ia de Investigaci\'on y Postgrado, Universidad de La Serena, 1700000, Chile.
   \and Universidad Andres Bello, Facultad de Ciencias Exactas, Departamento de F{\'i}sica y Astronom{\'i}a - Instituto de Astrof{\'i}sica, Autopista Concepci\'on-Talcahuano 7100, Talcahuano, Chile
   \and Gemini Observatory NSF's NOIRLab, 670 N. A'ohoku Place, Hilo, Hawaii, 96720, USA
   \and Rutgers the State University of New Jersey, 136 Frelinghuysen Ave., Piscataway, NJ 08854, USA
   \and Centro de Astronom\'ia (CITEVA), Universidad de Antofagasta, Av. Angamos 601, Antofagasta, Chile
   \and Millennium Institute of Astrophysics, Av. Vicuña Mackenna 4860, 782-0436 Macul, Santiago, Chile
   \and Instituto de Astrofísica, Pontificia Universidad Católica de Chile, Av. Vicuña Mackenna 4860, 782-0436 Macul, Santiago, Chile
   \and Instituto de Astrofísica, Departamento de Física y Astronomía, Facultad de Ciencias Exactas, Universidad Andres Bello, Fernandez Concha, 700, Las Condes, Santiago, Chile
}

   \date{Received / Accepted }

  \abstract
    {
    The characterization of Globular Clusters (GCs) in the Galactic bulge is a challenging task due to high extinction and severe stellar crowding. VVV-CL001 is a poorly studied GC located in the inner bulge, known for its extremely old age, extreme velocity, and low metallicity. Given its unique properties, a detailed study of this cluster can provide valuable insights into the early chemical and dynamical evolution of the Milky Way (MW).
    }
   {
    This study aims to derive the fundamental parameters of VVV-CL001 including metallicity, heliocentric velocity, proper motions, structural properties, orbit, and age in order to improve its origin and role in order to improve our understanding of its origin and role in the early evolution of the Milky Way.
    } 
   {
    We combined spectroscopic, astrometric, and photometric data to characterize VVV-CL001. Metallicity and radial velocity were determined from medium-resolution spectra obtained with FORS2 at the Very Large Telescope. Proper motions were derived using Gaia DR3 data. Near-infrared photometry from the FourStar instrument on Magellan was used to refine the cluster's position, construct a radial density profile, and estimate its age, distance, and reddening.
    }
   {
    Our results confirm that VVV-CL001 is an old ($12.1^{+1.0}_{-1.2}$ Gyr), metal-poor ([Fe/H] $= -2.25 \pm 0.05$~dex) globular cluster located at a heliocentric position of d$_\odot = 7.1^{+1.3}_{-1.1}$~kpc, with a reddening of E($J-K_\mathrm{s}$)  $= 1.40^{+0.01}_{-0.02}$. Its mean proper motions are $\mu_\alpha^* = -3.68 \pm 0.09$~mas~yr$^{-1}$ and $\mu_\delta = -1.76 \pm 0.10$~mas~yr$^{-1}$, and it exhibits a radial velocity of $-334 \pm 4$~km~s$^{-1}$. The cluster follows a retrograde-prograde, eccentric ($e = 0.76^{+0.10}_{-0.14}$) orbit, confined within the Galactic plane ($|Z|_{\rm max} = 1.0^{+0.45}_{-0.32}$~kpc) and inside the bar's radius of influence ($R < 5$~kpc), with a pericenter of $r_{\rm peri}= 0.6^{+0.3}_{-0.2}$ kpc and an apocenter of $r_{\rm apo}= 4.5^{+2.5}_{-1.2}$ kpc.
    }
    {
    These orbital properties, combined with its ancient age and low metallicity, strongly support an \textit{in-situ} origin for VVV-CL001 and likely member of the disk GC system that was captured by the potential of the bar during its formation. Thus, VVV-CL001 emerges as a fossil remnant of the earliest phases of Galactic assembly and a valuable tracer of the population that contributed to the formation of the inner thick disk and bulge, and part of the main progenitor of the MW - the inner disk and bulge. Our data distinguish it as  the main progenitor most metal-poor in-situ GC known. Our study highlights the relevance of detailed chemo-dynamical analyses in unveiling the origin of GBs in the inner Galaxy.
    }


    \authorrunning{Haro Moya et al.}
    \mail{w.haromoya@gmail.com}
    \titlerunning{Infrared photometry and CaT spectroscopy of VVV-CL001}
    \maketitle

\nolinenumbers
\section{Introduction} \label{sec:introduction}

GCs are dense, gravitationally bound systems of ten of thousands to millions of stars concentrated in a small volume. Their member stars formed simultaneously from the same molecular cloud, giving them a common age and metallicity. GCs are among the oldest structures in the Universe, with ages ranging from 10 to more than 13 Gyr \citep[e.g.,][]{ying2023,valcin2025age}. Virtually all of the older, more massive clusters display the multiple populations phenomenon. As such, they serve as important tracers of the formation and evolution of massive star clusters and indeed galaxies themselves  \citep{Gratton_2019,Beasley_2020}. The study of GCs provides key insights into the age, chemical enrichment history, kinematics, and dynamics of their host galaxies.

In the study of the formation and evolutionary history of the MW, GCs have played a prominent role. Through them, we have learned much about the history of the halo, discovering that its population is largely \textit{accreted} \citep{Massari_2019}, corroborating what simulations predict \citep{tumlinson_2009}. This has been possible thanks to the accessibility of this component with its null or low extinction, which has allowed the study of the GC population in the MW halo in great detail.

In contrast, simulations predict that the bulge and disk population of the MW are predominantly dominated by \textit{in-situ} population \citep{gargiulo_2019,bekki_2011}, making the GC population in these regions of great interest for understanding the \textit{in-situ} formation and the early evolution of the proto-MW, dubbed the main progenitor. However, optical observations of these MW components are strongly inhibited by the presence of dust and gas, which generates high crowding, high and often variable extinction over even the small angular sizes of a GC at many kpc  distance. Therefore, one must resort to observations at longer wavelengths that penetrate dust and gas, such as the near-infrared (NIR) \citep[see][]{Valenti_2007,Lim_2022,Cohen_2018}. In addition, crowding and field contamination also limit study of  these regions.

Over the past decades, extensive photometric and spectroscopic surveys such as the Two Micron All Sky Survey (2MASS; \cite{Skrutskie_2006}), the VISTA Variables in the Via Lactea Survey (VVV) and its extension, the VVVX Survey \citep{Minniti10,Saito2012,Smith_2018}, the Apache Point Observatory Galactic Evolution Experiment(APOGEE; \cite{majewski2017}), and the Gaia mission \citep{Gaia16,Gaia22}, have provided an unprecedented amount of data for Galactic objects, allowing for a much more comprehensive characterization of GCs. Additionally, observations from ground-based telescopes such as the Very Large Telescope (VLT) and GEMINI, along with space-based observatories like the Hubble Space Telescope (HST) and the James Webb Space Telescope (JWST), have significantly contributed to the study of these systems by offering high-resolution imaging and spectroscopy in optical and infrared wavelengths, to complement and augment the survey data. These studies have been instrumental in discovering and studying the structural and chemical properties of GCs. By combining astrometric, photometric, and spectroscopic information, researchers have refined the age-metallicity relation and calculated the orbital characteristics of most GCs, offering new insights into the assembly history of the Galaxy \citep{Garro_2021,Ceccarelli_2024,Woody_2021,Massari_2019,Belokurov_2023,Forbes_2010,Cohen_2021}.

In this context, VVV-CL001 was the first GC discovered with the VVV survey \citep{Minniti11}. Located in the Galactic bulge at coordinates $l = 5.27^\circ$, $b = 0.78^\circ$, it is projected very close to UKS1, a previously known GC, leading to speculation that both clusters might belong to the same system. \citet[][hereafter FT21]{Fernandez21} \defcitealias{Fernandez21}{FT21}, using APOGEE (Apache Point Observatory Galactic Evolution Experiment) high-resolution H-band spectra of two potential member stars and photometry from the VVV survey and 2MASS, determined that VVV-CL001 is located at a distance of $8.22^{+1.84}_{-1.93}$ kpc from the Sun, with an age of $11.9^{+3.12}_{-4.05}$ Gyr, a high radial velocity of $\sim -325$ km/s, and an extremely low metallicity of $[\mathrm{Fe}/\mathrm{H}] = -2.45 \pm 0.24$. They also derived an orbit, finding that it is halo-like and highly eccentric. \citet[][hereafter OC22]{Olivares22} \defcitealias{Olivares22}{OC22}, analyzing much lower-resolution optical and near-IR spectroscopic data from MUSE (Multi Unit Spectroscopic Explorer) for 55 stars in the field of the cluster and using photometry from VVV and VVVX, obtained similar values for the distance of $8.23 \pm 0.46$ kpc and radial velocity of $-324.9 \pm 0.8$ km/s but a substantially higher, but still very low, metallicity of $[\mathrm{Fe}/\mathrm{H}] = -2.04 \pm 0.02$. They also concluded it is basically confined to the bulge. These discrepancies in metallicity and orbit of the cluster highlight the need for more extensive spectroscopic studies to determine the true nature of VVV-CL001. Nevertheless, its extreme metallicity may point to an exotic early origin or unusual chemical enrichment conditions.

\citet{Belokurov_2023} introduced a novel method to classify GCs as either \textit{in-situ} or \textit{accreted}, based on their total energy ($E$) and the $z$-component of angular momentum ($L_z$), calibrated using [Al/Fe] abundance ratios. Applying this method to 165 GCs, they classified VVV-CL001 as an \textit{in-situ} object, agreeing with the orbital assessment of \citetalias{Olivares22} and in disagreement with the finding of \citetalias{Fernandez21}, who argued that the high orbital eccentricity of VVV-CL001 might instead favor a \textit{halo intruder} and therefore accreted origin. This \textit{in-situ} classification is further supported by the recent CAPOS survey \citep{geisler25}, which analyzed bulge GCs and included VVV-CL001 in their \textit{in-situ} sample based on chemodynamical criteria. However, this contrasts sharply with the cosmological simulations of \citet{boldrini2025new}, who identified VVV-CL001 as \textit{ex-situ} using a new energy threshold ($E < 0.7E_{\rm circ}(r^{*}_{\rm hm})$). Such discrepancies underscore the challenges in disentangling the origins of GCs, particularly those in the bulge/halo transition region, where dynamical and chemical signatures often overlap.

All previous photometric analyses of VVV-CL001 have been based on Red-Giant Branch (RGB) stars, which are not ideal for precise age determinations since, the RGB locus is not very sensitive to age. This insensitivity arises because the RGB phase, while long-lived, converges stars of different initial masses onto a similar path in the color-magnitude diagram (CMD). Thus, the age of VVV-CL001 remains uncertain and needs to be determined using alternative methods. Deep photometric studies of the cluster can provide insight into evolutionary phases such as the Sub-Giant Branch (SGB) and the main-sequence turn-off (MSTO), which are more suitable for age determination. Crucially, although the SGB phase is longer than the RGB phase, for a given star, the position of stars on the SGB in the CMD is highly sensitive to their initial mass and thus the cluster's age. This sensitivity, combined with the shorter evolutionary timescales and larger separation between isochrones at the MSTO, makes these phases superior chronometers. Although current data for VVV-CL001, due to instrumental limitations, do not reach the MSTO, which is critical for deriving the most reliable age, incorporating SGB stars into the analysis represents a significant improvement over previous methods that rely solely on RGB stars.

In this work, we present a detailed analysis of the GC VVV-CL001 using spectroscopic data from the FORS2 (Focal Reducer and Low Dispersion Spectrograph 2) instrument on the VLT (Very Large Telescope), astrometric data from Gaia DR3 (Gaia Data Release 3), and photometric data from the VVV survey, 2MASS, and the FourStar instrument on the Magellan Baade Telescope. Our main objectives are to determine the cluster metallicity and age using both SGB and RGB stars and to study the cluster's orbit within the Galaxy. The paper is organized as follows: Section \ref{s_data} describes the data and data reduction processes; Section \ref{s_spec} presents the methods used to determine the cluster's radial velocity and metallicity; Section \ref{s_phot} discusses the analysis of IR photometric data, including the determination of the cluster's center, density profile, proper motions, CMD, isochrone fitting; Section \ref{s_orbit} explores the methodology used to determine the cluster's Galactic orbit; and finally, Section \ref{Dis_concl} presents our discussion and conclusions.

\section{Observations and data reduction}
\label{s_data}

\subsection{Photometric data}
\label{ss_dataphot}
\begin{figure}
    \centering
    \includegraphics[width=.9\columnwidth]{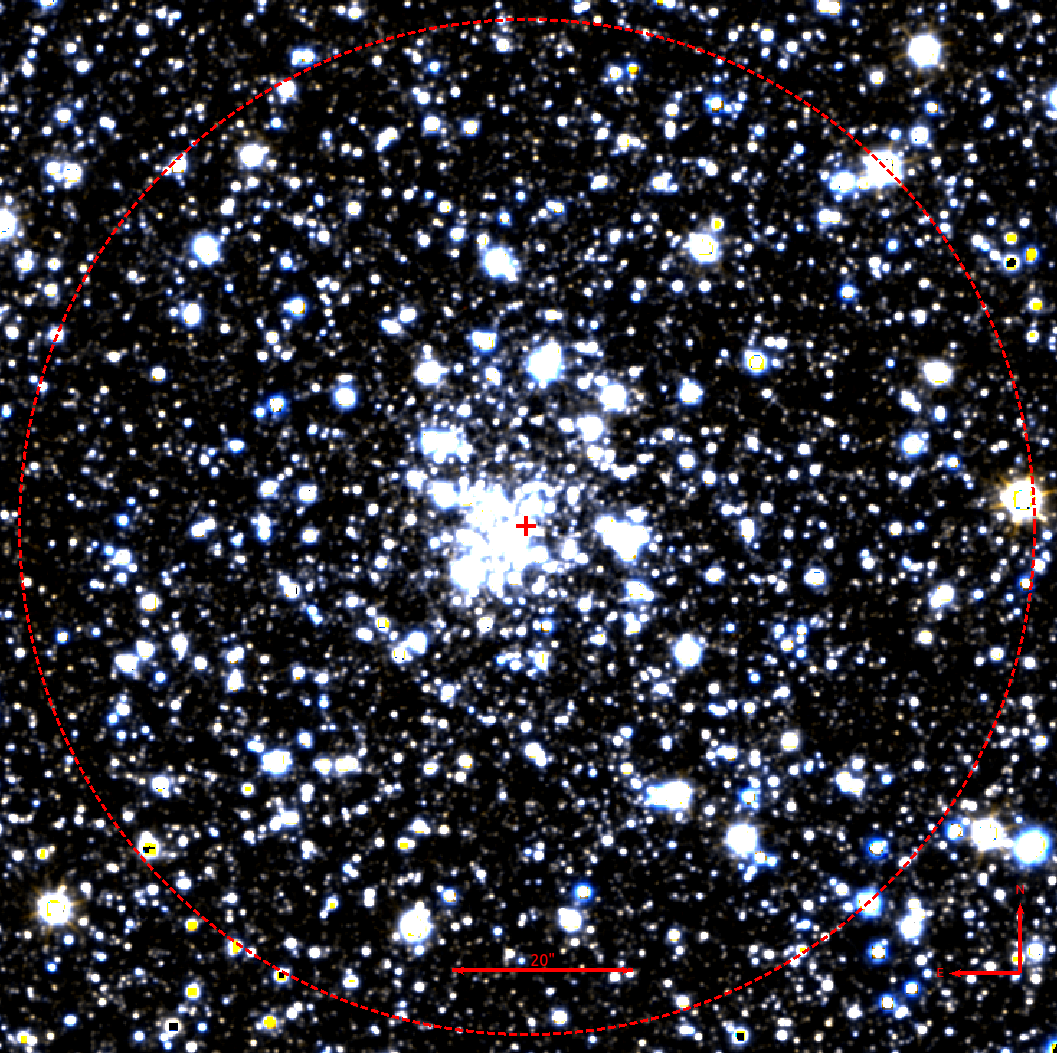}
    \caption{False-color composite image using the infrared J (blue) and Ks (red) bands, centered on the GC VVV-CL001. The red dashed circle marks the core radius ($r_c=0\farcm94$), while the cross indicates the cluster's central position.}
    \label{fig:VVV-CL001}
\end{figure}

Deep NIR photometric data were collected at Las Campanas Observatory during one night of observation on 2012 May 9, with the FourStar camera at the 6.5m Baade telescope (program ID CN2012A-034, PI: S. Villanova). Five frames were collected in both the $J$ and $K_\mathrm{s}$ bands, each one divided in four jittered sub-exposures, for a total exposure time of 38$^s$ ($J$ band) and 6$^s$ ($K_\mathrm{s}$) for each frame. The target cluster was not positioned at the center of the resulting $5\farcm7\times5\farcm7$ field to avoid the gaps between the four detectors. As a consequence, our data are limited to $3\arcmin$ from the cluster center in the West direction (see Fig. \ref{fig:VVV-CL001})

The data were reduced using standard IRAF\footnote{IRAF is distributed by the National Optical Astronomy Observatories, which are operated by the Association of Universities for Research in Astronomy, Inc., under cooperative agreement with the National Science Foundation.} routines. The PSF photometry was performed with the VVV-SkZ\_pipeline \citep{Mauro13}, based on the DAOPHOT~II and ALLFRAME codes \citep{Stetson94}. The results were calibrated on the 2MASS astrometric and photometric system, as detailed in \citet{Moni11} and \citet{Chene12}. The catalog was then cleaned of spurious detections, mainly found at redder colors around $K_\mathrm{s}\approx16$, caused by defects of the detector and heavily saturated stars. Our FourStar photometry resulted more than two magnitudes deeper than VVV, but it lacked stars brighter than $K_\mathrm{s}=12.6$ due to saturation. We therefore obtained a shallower photometric catalog running the same pipeline on the VVV frames, and calibrated it on the 2MASS astrometric and photometric system. Then, we cross-matched the resulting catalogs, and added to our data the 2MASS and VVV sources undetected by FourStars, mainly bright objects up to $K_\mathrm{s}=3$. We thus eventually obtained our final photometric catalog, that includes 234,240 objects.

\subsection{VVV proper motions}
\label{ss_dataPMs}

\begin{figure}
\includegraphics[width=\columnwidth]{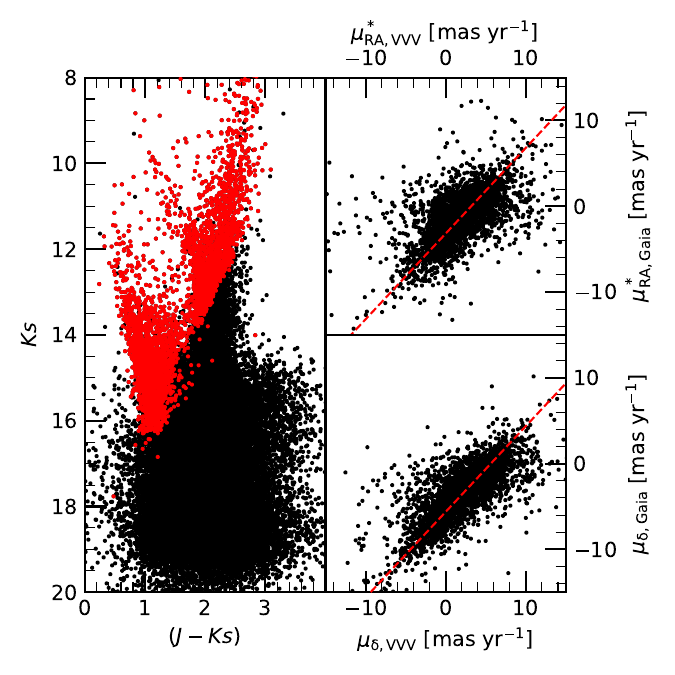}
\caption{{\it Left panel}: CMD of the VVV sources within 5$\arcmin$ of the cluster center with (red dots) and without (black dots) proper motion from Gaia catalog. {\it Right panel}: comparison of proper motions in galactic coordinates from Gaia and VVV. The red line indicates the fit of the data with a linear relation of unit slope.}
\label{f_Gaiacomp}
\end{figure}

The large interstellar reddening along the line of sight (\citet{Minniti11} estimate $\mathrm{E(B-V)>2.0}$) strongly limits the use of Gaia data, which are based on observations at visible wavelengths. In fact, the Gaia DR3 catalog \citep{Gaia22} reports proper motions (PMs) only for the bluest and brightest stars of the field, as shown in the left panel of Fig.~\ref{f_Gaiacomp}. This covers only the brightest section of the cluster RGB, which is also scarcely populated, hence only a few cluster stars are expected to be included in the catalog. We therefore measured PMs from 83 VVV epochs taken between 2010 and 2015.
The VVV PMs were measured with the procedure described in detail in \citet{Contreras17}, which returns values relative to the mean motion of bulge RGB stars in Galactic coordinates $(\mu_l^*,\mu_b)$, where $\mu_l^*=\mu_l\cdot\cos b$. We transformed these Galactic components to the equatorial ones $(\mu_\alpha^*,\mu_\delta)$, where $\mu_\alpha^*=\mu_\alpha\cdot\cos\delta$, by means of the formulae of \citet{Poleski13}. The zero-point offset of the VVV relative PMs was determined from the axis intercept of the Gaia-vs-VVV relation for the $\approx4200$ stars in common, which was fit with a linear relation of unit slope, as shown in the right panels of Fig. \ref{f_Gaiacomp}. VVV frames saturated at $K_\mathrm{s}=12.5$, and the VVV PMs were eventually complemented with Gaia values for the brighter sources. The final PM catalog included $\approx48700$ VVV sources down to $K_\mathrm{s}=18.5$ and within 5$\arcmin$ from the cluster center.

\subsection{Spectroscopic data}
\label{ss_dataspec}

\begin{figure}
\includegraphics[width=.9\columnwidth]{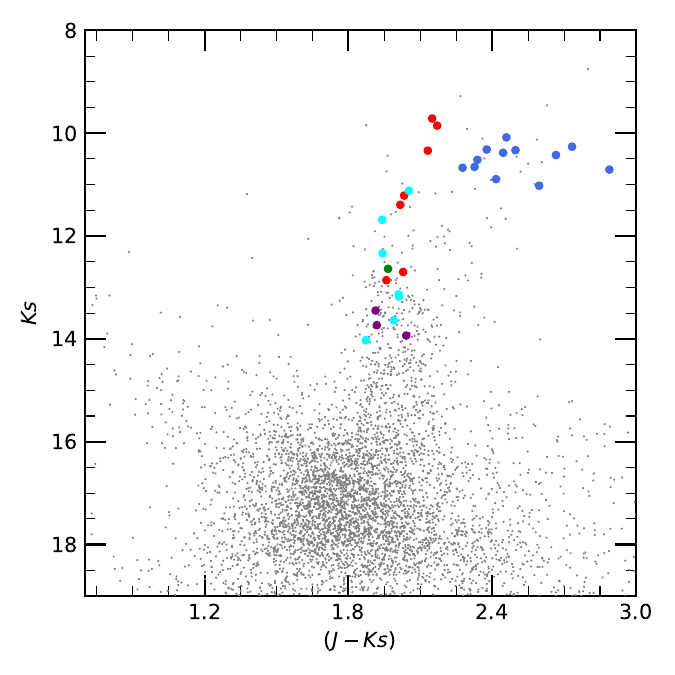}
\caption{Decontaminated VVV CMD of VVV-CL001 . All sources detected within 1$\arcmin$ from the cluster center are shown with gray dots. Spectroscopic targets are marked with large colored points: blue, cyan,  green and purple symbols are stars rejected as cluster members because they have a discrepant distance to the cluster center, mean RV, metallicity and PMs, respectively. Red points represent spectroscopically confirmed cluster members.}
\label{f:CL001_cmd}
\end{figure}

Using the instrument FORS2 on the Very Large Telescope (Paranal, Chile), we obtained spectra of red giant stars belonging to the cluster VVV-CL001 and its surrounding field. Targets were selected according to their position in the cluster CMDs built from VVV photometry (Fig. \ref{f:CL001_cmd}). Observations were performed as part of the program 089.D-0392(PI: D Geisler) in service mode with the mask exchange unit (MXU), with the 1028z+29 grism and OG590+32 filter. The cluster was located in the master CCD and the secondary CCD was used for the observation of field stars. Slits located in the master and slave chips have a size of 1$\arcsec$ wide and $4-8\arcsec$ long, and pixels were binned 2$\times$2, yielding a plate scale of $0\farcs25$ pixel$^{-1}$ and a dispersion of $\sim$ 0.85 \AA \space pixel$^{-1}$. Resulting spectra are centered in the Ca II triplet region ($\sim8600$~\AA) covering a spectral range from 7750 to 9500 \AA. We obtained 7 exposures of 660s each in order to reach an adequate final S/N. In this paper we only present the analysis of stars located in the master chip. Field star analysis will be included in an upcoming paper.  We further note that these data were obtained during the same program as the other data reported in \citet[][hereafter G23]{geisler23} \defcitealias{geisler23}{G23}.

The pipeline provided by ESO (version 2.8) was used to perform the bias, flatfield, distortion correction and the wavelength calibration.
The necessary calibration images were acquired by the ESO staff. We performed the extraction and the sky subtraction using the task {\it apall}  of IRAF. IRAF was also used for the combination of the spectra ({\it scombine} task) and the normalization of the combined spectra ({\it continuum} task).

\section{CaT spectroscopy}
\label{s_spec}

\subsection{Heliocentric radial velocity and equivalent width measurements}
\label{ss_rv}
We measured the  heliocentric radial velocities (RV) and the equivalent widths (EW) of our targets following the method described in detail in \citetalias{geisler23}. Briefly, we cross-correlated the observed spectra with template spectra, acquired with the same telescope and instrument, of stars belonging to Galactic open and GCs \citep{col04} using the IRAF task {\it fxcor}. We adopt as the final RV the average of those cross-correlation results with a standard deviation of $\sim$ 6~km~s$^{-1}$. This standard deviation is added in quadrature to the error corresponding to the misalignment of the star in the slit (4.5~km~s$^{-1}$) to obtain the final adopted error for the RVs (7.5~km~s$^{-1}$). \citet{par09} explains the procedure to correct for the effect introduced by the offset between the star and slit centers.

We used a combination of a Gaussian and a Lorentzian function \citep{col04} and the bandpasses from \citet{vas15} to measure the equivalent widths (EW) of the CaT lines on the normalized spectra. The estimated errors in the EW measurements are between $\sim$ 0.1 $-$ 0.5 \AA.  As described in the next section, we followed the idea of \citetalias{geisler23} and calculated the cluster metallicity using three different calibrations: \citet{vas15}, \citet{vas18} and \citet{DP20}, hereafter V15, V18 and DP20, respectively. \defcitealias{vas15}{V15} \defcitealias{vas18}{V18} \defcitealias{DP20}{DP20} The sum of the EW of the CaT lines ($\Sigma$EW) correlates with the metallicity of the clusters \citep{arz88}, therefore the construction of this index is an important step. \citetalias{vas15} and \citetalias{vas18} are on the same scale as \citet{sav12} and use only the two strongest CaT lines to calculate the $\Sigma$EW. In the case of \citetalias{DP20}, all three CaT lines are used. The $\Sigma$EW from \citetalias{DP20} are based on the EWs measurements from \citet{par09,par15} so they are on the scale of \citet{col04}. In order to be completely consistent with these three calibrations, we calculate the $\Sigma$EW using both the two strongest lines as well as all three lines. \citet{sav12}, \citetalias{vas15} and \citetalias{vas18} showed that small differences exist between their EW measurements for the same spectrum even when the same pseudo-continuum and bandpass regions are adopted. Therefore, for the particular case of the calibration of \citetalias{vas18} it is necessary to evaluate if such a difference is present between our EW measurements and those from \citetalias{vas18}. In \citetalias{geisler23} we used 119 spectra in 9 Bulge Globular Clusters (BGC) from \citetalias{vas18} to perform this comparison. \citetalias{geisler23} found the following correlation between the sum of the EW of the two strongest CaT lines between \citetalias{vas18} and our work (OW):
\begin{equation}
    \sum{\rm EW_{V18}}=0.95\cdot \sum{\rm EW_{OW}} + 0.06.
\label{eq:EW}
\end{equation}
We then corrected our $\Sigma$EW from the two strongest lines according to equation \ref{eq:EW}. As is shown in \citetalias{geisler23} this EW offset between both studies implies a small error in the metallicity of only about 0.03 dex.

\subsection{Metallicity determination and membership}
\label{ss_met}

\begin{figure}
\includegraphics[width=.9\columnwidth]{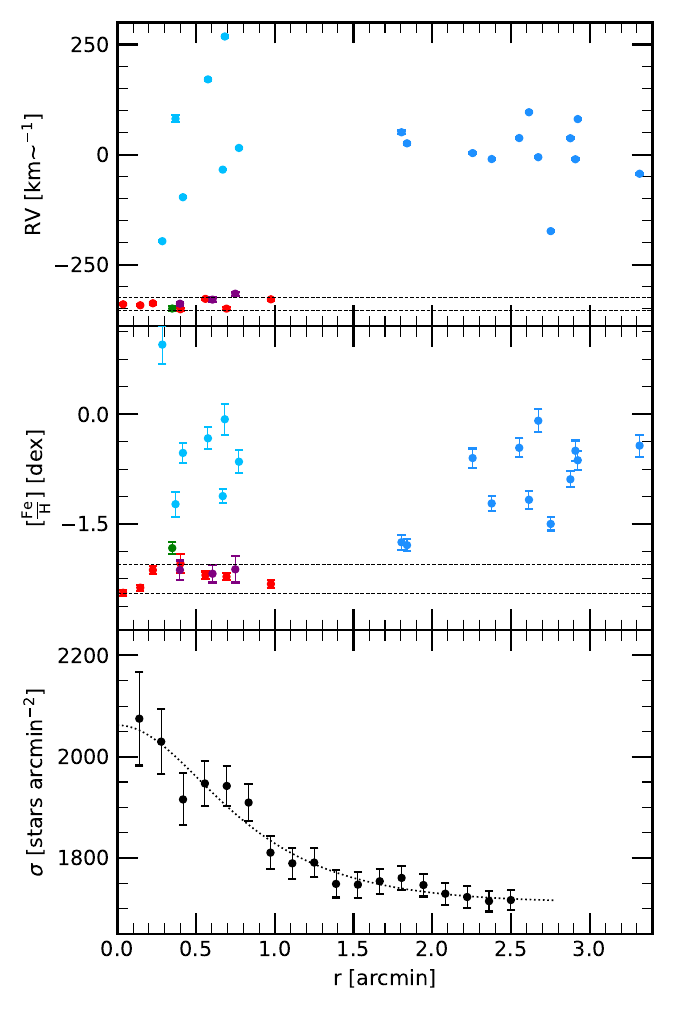}
\caption{
{\it Upper panel}:
RV of the spectroscopic targets as a function of distance from the cluster center. The horizontal lines represent our velocity  cuts ($\pm$ 15 km s$^{-1}$).
{\it Middle panel}:
metallicity of the spectroscopic targets as a function of distance from the cluster center. The horizontal lines represent our metallicity cuts ($\pm$ 0.20 dex). For both panels the color code is the same as in Fig. \ref{f:CL001_cmd}.
{\it Lower panel}: radial density profile of the photometric sources with respect to the cluster center. The dotted curve indicates the best fit of a \citet{King62} radial profile.}
\label{f_radall}
\end{figure}

The $\Sigma$EW depends not only on the metallicity, but also on the effective temperature and surface gravity \citep{ard91,ols91}. These two latter dependencies can be removed by the so-called reduced EW ($W'$), which uses the dependence of these quantities along the upper giant branch and corrects for the difference of magnitude between the observed star and the horizontal branch in a particular filter.
Much work has been carried out to calibrate  $\Sigma$EW with  metallicity using a variety of filters (see \citetalias{DP20} for a summary and compilation). In our case we use the $K_s$ magnitudes from the VVV survey \citep{Saito2012} so $W'$ can be expressed as follows:
\begin{equation}
    W' = \Sigma EW + \beta_{K_s}(K_s-K_{s,HB}).
    \label{eq:reduced_ew}
\end{equation}
We used the corresponding $\beta_{K_s}$ for the corresponding calibration: $\beta_{K_s}$ = 0.384 \citepalias{vas15}, $\beta_{K_s}$ = 0.55 \citepalias{vas18} and $\beta_{K_s}$ = 0.37 \citepalias{DP20}. To calculate  $K_{s,HB}$ we followed the procedure proposed by \citet{mau14} which is explained in detail in Section~5 of \citetalias{geisler23}. 

The three calibrations between [Fe/H] and $W'$ used are:
\begin{equation}
    \mathrm{[Fe/H]_{V15}} = -3.150 + 0.432W' + 0.006W'^2,
    \label{eq:met_V15}
\end{equation}
\begin{equation}
    \mathrm{[Fe/H]_{V18}} = -2.68 + 0.13W' + 0.055W'^2,
     \label{eq:met_V18}
\end{equation}
\begin{equation}
    \mathrm{[Fe/H]_{DP20}} = -2.917 + 0.353W'.
     \label{eq:met_DP20}
\end{equation}
We then calculated each target metallicity using the three calibrations, with a mean error of $\sim$ 0.20 dex. Following \citetalias{geisler23} we used the calibrations of \citetalias{vas15}, \citetalias{vas18} and \citetalias{DP20} for comparison purpose, but our preferred calibration is \citetalias{vas18} because it is based on the metallicity scale of \citet{dias+16a,dias+16b}. That scale includes MW GCs covering all the metallicity range of the bulge clusters. Therefore, in the subsequent analysis we use for our targets the metallicity values derived from the calibration of \citetalias{vas18}.

The applied discrimination method between cluster members and surrounding field stars is the same used by our group in previous work \citep[][]{par09,par15,dias+21,dias+22,debortoli+22,geisler23}. We refer the reader to these papers for a detailed description  but, briefly, the method discards as possible cluster members those targets that fall outside of certain cuts in distance from the mean cluster center, RV and [Fe/H]. We adopted RV cuts of $\pm$ 15 km s$^{-1}$ and metallicity cuts of $\pm$0.20~dex, which are typical errors for individual stars. The cluster core radius ($r_c = 0\farcm94$) was adopted as the distance cut, in order to maximize the probability that the selected stars are members of the cluster (see section \ref{ss_photdensity} and Fig. \ref{f_radall}).

\begin{figure}
\includegraphics[width=.9\columnwidth]{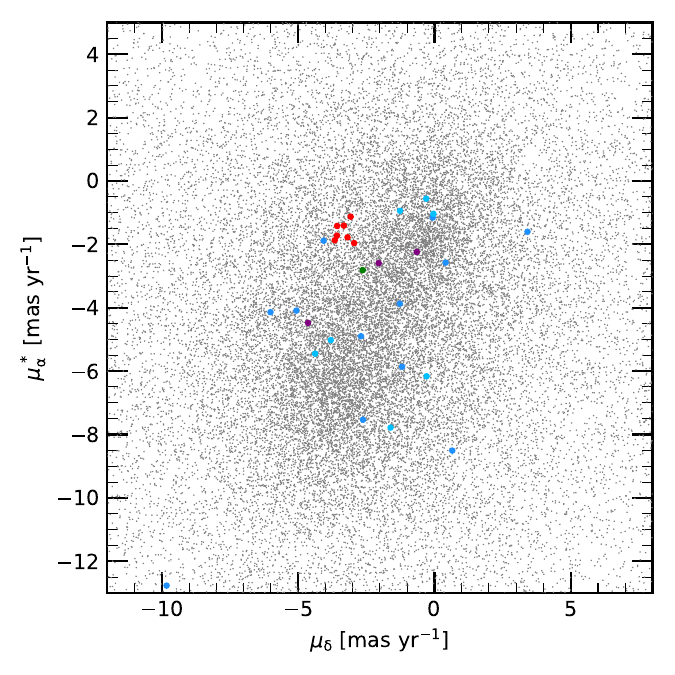}
\caption{PM plane for cluster VVV-CL001. Gray points stand for stars from the Gaia DR3 catalogue and large colored points represent our spectroscopic targets. Color code is the same as in Fig. \ref{f:CL001_cmd} .}
\label{f_PM}
\end{figure}

The behavior of RV and metallicity, respectively, with distance from the cluster center are shown in Fig. \ref{f_radall}. Fig. \ref{f_PM} shows the proper motion plane for CL001 where targets (large colored points) are plotted jointly with stars from the Gaia DR3 catalog centered on the cluster coordinates.  In these plots we represent targets with the following color code (the same used in our previous work  e.g. \citetalias{geisler23}): blue symbols represent field stars located at a distance from the cluster center larger than the adopted radius, cyan and green are stars with RV or  metallicity, respectively, outside the adopted cuts. We finally check that the stars that have passed the three aforementioned cuts also have similar proper motions (Fig. \ref{f_PM}). Targets rejected because of their discrepant proper motions are marked in purple. Finally, the seven stars that passed all the membership criteria and, therefore are considered cluster members, are plotted in red. 

For cluster members we include in Table \ref{t:individuals} the star identification, the equatorial coordinates, RV, $K_s-K_{s,HB}$, the sum of the measured EWs of the CaT line ($\Sigma$EW$_{2L}$ for two lines and $\Sigma$EW$_{3L}$ for three lines), the \citetalias{vas18} metallicity and proper motions. Using the information of this table and equations \ref{eq:met_V15} and \ref{eq:met_DP20} individual metallicities using \citetalias{vas15} and \citetalias{DP20} calibrations can be straightforwardly calculated.

Using cluster member stars, we calculate mean values of [Fe/H]$_{V15}$ = -2.31 $\pm$ 0.08 dex , [Fe/H]$_{V18}$ =  -2.25 $\pm$ 0.05 dex , [Fe/H]$_{DP20}$ = -2.13 $\pm$ 0.09 dex, RV = -339.7$\pm$ 3.5 km$^{-1}$, $\mu^*_{\alpha}$ = -3.33 $\pm$ 0.10 and $\mu_{\delta}$ = -1.61 $\pm$ 0.11  for cluster mean RVs, metallicity and PM, respectively. The quoted errors correspond to the standard error of the mean. As we found in \citetalias{geisler23} for a sample of 12 BGCs, the mean metallicities corresponding to the \citetalias{vas15} and \citetalias{DP20} calibrations maintain good agreement with the values given by the \citetalias{vas18} calibration.

Our sample has some stars in common with previous spectroscopic works. In particular, 5 cluster member stars in our sample have been previously observed with MUSE \citepalias{Olivares22} and 2 with APOGEE \citepalias{Fernandez21}. Table \ref{t:met_comp} includes a comparison between the metallicity values obtained by the different works. As can be seen, the metallicity values derived in this work show a very good agreement with those derived by \citetalias{Fernandez21} for the 2 stars in common. However, the values from \citetalias{Olivares22} present a mean offset of 0.27 dex with respect to our determinations.

\begin{table*}
\caption{Measured Values for Member Stars}             
\label{t:individuals}      
\centering                 
\footnotesize
\begin{tabular}{lccccccccc}
\hline
\multicolumn{1}{c}{Star ID} & 
\multicolumn{1}{c}{R.A.} & 
\multicolumn{1}{c}{Dec} & 
\multicolumn{1}{c}{RV} & 
\multicolumn{1}{c}{$K_s-K_{s,HB}$} & 
\multicolumn{1}{c}{$\Sigma$EW$_{2L}$} &
\multicolumn{1}{c}{$\Sigma$EW$_{3L}$} &
\multicolumn{1}{c}{[Fe/H]$_{V18}$} &
\multicolumn{1}{c}{$\mu^*_{\alpha}$} &
\multicolumn{1}{c}{$\mu_{\delta}$}\\
\multicolumn{1}{c}{} &
\multicolumn{1}{c}{J2000 [$^o$]} & 
\multicolumn{1}{c}{J2000 [$^o$]}  & 
\multicolumn{1}{c}{km s$^{-1}$}  & 
\multicolumn{1}{c}{mag} & 
\multicolumn{1}{c}{\AA} & 
\multicolumn{1}{c}{\AA} & 
\multicolumn{1}{c}{dex} &
\multicolumn{1}{c}{mas yr$^{-1}$} &
\multicolumn{1}{c}{mas yr$^{-1}$} \\
\hline
\noalign{\smallskip}
400169332 & 17.9124 & $-$24.0275 & $-328.8\pm3.6$ & $-3.58\pm0.15$ & $3.06\pm0.11$ & $3.60\pm0.12$ & $-2.32\pm0.05$ & $-3.57\pm0.88$ & $-1.42\pm0.58$ \\
400150514 & 17.9114 & $-$24.0184 & $-351.3\pm3.4$ & $-5.95\pm0.15$ & $4.82\pm0.31$ & $6.29\pm0.36$ & $-2.04\pm0.13$ & $-3.32\pm0.21$ & $-1.41\pm0.16$ \\
400159772 & 17.9119 & $-$24.0159 & $-342.3\pm2.6$ & $-6.54\pm0.16$ & $4.03\pm0.10$ & $4.89\pm0.12$ & $-2.37\pm0.04$ & $-3.07\pm0.21$ & $-1.13\pm0.13$ \\
400157893 & 17.9118 & $-$24.0149 & $-340.1\pm2.5$ & $-6.62\pm0.15$ & $3.79\pm0.08$ & $4.53\pm0.09$ & $-2.44\pm0.04$ & $-3.66\pm0.19$ & $-1.87\pm0.12$ \\
400160612 & 17.9119 & $-$24.0119 & $-337.7\pm3.3$ & $-4.91\pm0.15$ & $4.16\pm0.07$ & $4.86\pm0.08$ & $-2.13\pm0.05$ & $-3.19\pm0.30$ & $-1.78\pm0.20$ \\
400147752 & 17.9112 & $-$24.0094 & $-327.7\pm2.7$ & $-5.07\pm0.15$ & $4.02\pm0.09$ & $4.66\pm0.10$ & $-2.20\pm0.05$ & $-3.57\pm0.25$ & $-1.72\pm0.17$ \\
400151668 & 17.9114 & $-$24.0039 & $-349.9\pm3.5$ & $-3.46\pm0.15$ & $3.32\pm0.11$ & $4.11\pm0.12$ & $-2.22\pm0.05$ & $-2.95\pm0.90$ & $-1.96\pm0.58$ \\

\noalign{\smallskip} 
\hline\end{tabular}
\end{table*}

\begin{table}
\caption{Comparison with previous spectroscopic metallicity determinations}
\label{t:met_comp} 
\centering         
\footnotesize
\begin{tabular}{lccc}
\hline
\multicolumn{1}{c}{Star ID} & 
\multicolumn{1}{c}{[Fe/H]$_{V18}$} & 
\multicolumn{1}{c}{[Fe/H]$_{MUSE}$} & 
\multicolumn{1}{c}{[Fe/H]$_{APOGEE}$} \\

\noalign{\smallskip}
 & dex & dex    & dex    \\
\hline
400169332 & $-$2.32 & --    & --    \\
400150514 & $-$2.04 & $-$1.86 & --     \\
400159772 & $-$2.37 & $-$2.04 & $-$2.44 \\
400157893 & $-$2.44 & $-$2.09 & $-$2.47 \\
400160612 & $-$2.13 & $-$1.90  & --      \\
400147752 & $-$2.20 & $-$1.94 & --   \\
400151668 & $-$2.22 & --    & --   \\
\noalign{\smallskip} 
\hline\end{tabular}

\end{table}
\section{Infrared photometry}
\label{s_phot}

\subsection{Cluster center and density profile}
\label{ss_photdensity}
To determine the center of the cluster, we first calculated the average position of all the sources detected within $0\farcm5$ from the center given by \citet{Minniti11}. The result was used as the new center and the procedure iterated until convergence. The position of the center changed by less than $0\farcs2$ in each iteration, and eventually converged to the coordinates (RA, dec)=$(17^h 54^m 42\fs22, -24\degr 00\arcmin 52\farcs3)$, only $1\farcs8$ away from the original center of \citet{Minniti11}.

The cluster density profile was obtained by measuring the stellar density in concentric annuli of width $\Delta r=0\farcm14$, centered at the coordinates derived above. We also considered different binning schemes, including wider, narrower and/or overlapping annuli, finding no substantial difference in the results. The profile was not extended beyond $2\farcm5$ from the center, where the mosaic nature of our photometric catalog caused a sudden discontinuity of the source density in the north-western direction. The results are shown in the lower panel of Fig. \ref{f_radall}, where the vertical bars indicate the Poissonian errors of stellar counts. In the Figure we also show a tentative fit of the data with a \citet{King62} profile, which returns $r_c=0\farcm94$ and $r_t=3\farcm48$ for the core and tidal radius, respectively. These results suggest that VVV-CL001 is a small ($r_t=7.6$~pc at a distance of 7.5~kpc) and sparse (concentration parameter $c=0.57$) object, very similar to the smallest Galactic GCs such as Whiting~1 \citep[c=0.55,][]{Carraro2007}, FSR1735 \citep[c=0.56,][]{CarballoBello2016}, and Palomar~10 \citep[c=0.58,][]{Kaisler1997}. However, we consider this result only indicative and not fully reliable because a King profile should be inappropriate to describe a small cluster embedded in the gravitationally complex inner Galactic regions. Deviations from the King profile should be expected, in particular at intermediate radii between the innermost and the external regions, which tend to push the fit profile to higher $r_t$ and $r_c$ and lower $c$, as previously observed in similar situations \citep[e.g.,][]{Moni11,CarballoBello2016}.

\subsection{Proper motions}
\label{ss_pm}
Fig. \ref{f_PM} shows the PMs of all the Gaia-VVV sources within 5$\arcmin$ from the cluster center. Two main populations can be identified, with approximately elliptical distributions around the points $(\mu_\alpha^*,\mu_\delta)\approx(-3.5, -6.0) ~\mathrm{mas~yr^{-1}}$ and $\approx(-0.5, -2.0) ~\mathrm{mas~yr^{-1}}$. An inspection of the CMD reveals that the first population is clearly more distant and redder, and it lacks a noticeable group of young blue stars. These characteristics identify it as the bulge main stellar population. The second group, on the other hand, is rich in blue main-sequence stars, corresponding to the foreground disk.

In the same Figure, we also show the proper motions of the spectroscopic targets, coded with the same colors as Fig.~\ref{f_radall}. The stars identified as cluster members (red dots) clearly reveal the cluster. However, the small cluster is embedded in a crowded area, and its proper motion is not too different from the field. As a consequence, no overdensity of stars can be identified at this position in the general PM catalog, even if restricted to stars within 1$\arcmin$ from the center. We therefore cleaned the diagram from the field contamination, with a statistical algorithm similar to that employed later in the CMD (Sect.~\ref{ss_cmd}), but applied to the PM plane. In brief, we created a cluster catalog with all the sources within 1$\arcmin$ from the cluster center, and a comparison field catalog with the stars in an annulus of equal area and inner radius of 2$\arcmin$. This selection was made based on the density profile shown in Fig. \ref{f_radall}. Then, the closest source in the field catalog was identified for each object in the cluster area, and if the distance was lower than 1~mas~yr$^{-1}$, both stars were removed from their respective catalogs. The objects of the cluster area that survived this procedure formed our final cluster catalog cleaned from field contamination, and they are indicated in Fig.~\ref{f_PM} as color dots. Apart from some scarce and scattered residual contamination, the position of the cluster thus can be properly identified. We derived the cluster PM as the weighted average of the clumped stars in the de-contaminated list, to which we added the spectroscopically confirmed cluster members, finding $(\mu_\alpha^*,\mu_\delta)=(-3.68\pm0.09, -1.72\pm0.12) ~\mathrm{mas~yr^{-1}}$. Our results agree within approximately 1$\sigma$ with those of \citet{Vasiliev21} and \citetalias{Fernandez21}, while we have irreconcilable differences in the component $\mu_\delta$ with \citetalias{Olivares22}, as summarized in Table \ref{t_PM}.
\begin{table}[t]
\begin{center}
\caption{Cluster proper motions from the literature}
\label{t_PM}
\begin{tabular}{l l l}
\hline
\hline
Reference & \multicolumn{1}{c}{$\mu_\alpha^*$} & \multicolumn{1}{c}{\textbf{$\mu_\delta$}} \\
 & \multicolumn{1}{c}{mas~yr$^{-1}$} & \multicolumn{1}{c}{mas~yr$^{-1}$} \\
\hline
 \citetalias{Fernandez21} & $-3.41\pm0.5$ & $-1.97\pm0.5$ \\
\citet{Vasiliev21} & $-3.49\pm0.16$ & $-1.68\pm0.11$ \\
 \citetalias{Olivares22} & $-3.42\pm0.42$ & $-3.58\pm0.20$ \\
This work & $-3.68\pm0.09$ & $-1.76\pm0.10$  \\
\hline
\end{tabular}
\end{center}
\end{table}

\subsection{Color-Magnitude diagram}
\label{ss_cmd}

We cleaned the CMD from field contamination following the procedure of \citet{Moni11} and \citet{Moni21}, based on the method described by \citet{Gallart03}. For each star in the field region, the algorithm identifies the nearest star in the cluster region in terms of a distance metric $d$ in the CMD, defined as:
\begin{equation}
d=\sqrt{(\Delta K_\mathrm{s})^2+(k\times (\Delta (J-K_\mathrm{s}))^2},
\label{e_metric}
\end{equation}
where $k$ is an arbitrary coefficient that weights the color differences relative to magnitude differences. If $d$ is smaller than a given threshold ($d_\mathrm{max}$), the star is removed from the cluster catalog. For this study, we adopted $k$=2 and $d_\mathrm{max}=0.3$ \citep[see][for a discussion on these parameters]{Moni21}. The cluster region is defined as a circular area with a radius of $r_c = 0\farcm94$ around the estimated center (Section~\ref{ss_photdensity}), while the field region is selected as an annular region with an inner radius of $r_\mathrm{in} = 1\farcm2$, chosen so that its area matches that of the cluster region.

Any statistical decontamination procedure is based on the underlying hypothesis that the photometric properties of the contaminating field stars are identical to those in the control area. Large reddening variations in the field could violate this hypothesis, and cause the removal of the wrong stars in the cluster area, thus spoiling the results. We performed a series of decontamination tests to verify this possibility and to ensure the robustness of our results. We first selected eight circular regions of radius $0\farcm4$, located at a distance of $1\farcm5$ from the estimated cluster center, and we ran the decontamination procedure on each of them using as field area the circular region on the opposite side with respect to the cluster. This test revealed small but non-negligible residuals in the decontaminated CMD, due to the presence of a reddening gradient in the direction of the Galactic center, which is only 5$\fdg$5 away. However, the residuals disappeared when we decontaminated an annular region using another annular area with larger inner radius as the control field. This showed that the annular symmetry is enough to compensate for the smooth variations of reddening in the field, and the cluster decontamination is therefore reliable.

\begin{figure}
\centering
\includegraphics[width=.9\columnwidth]{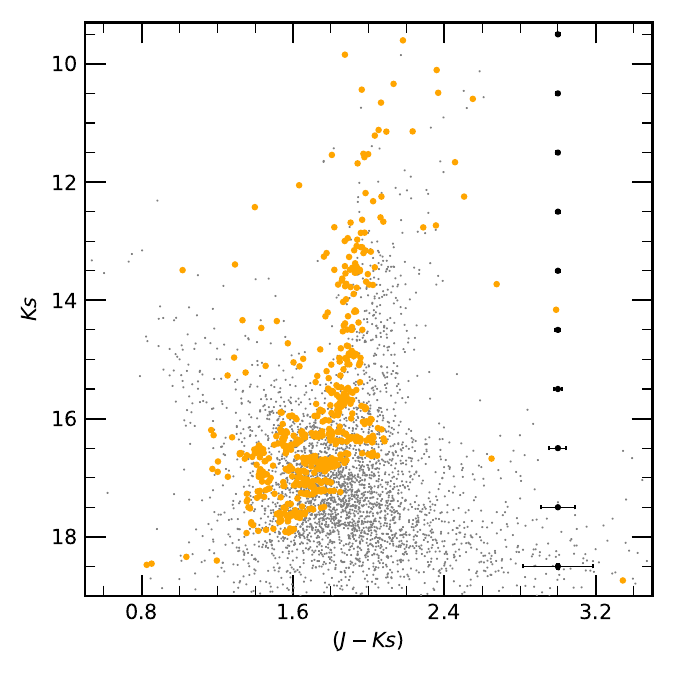}
\caption{CMD of the photometry obtained in the VVV CL001 field  within a radius of  $0\farcm94$ from the center of the cluster (gray points). Orange points, the stars resulting from the statistical decontamination process. Black symbols, the average photometric errors per magnitude bin.}
\label{f_cmd}
\end{figure}

Fig. \ref{f_cmd} shows the result of the statistical decontamination process with the orange points. The RGB is clearly visible at magnitudes brighter than $\sim16$. At fainter magnitudes, the SGB is less evident due to larger photometric errors and some residual contaminating stars. 

\subsection{Reddening and age determination}\label{ssec:age}

\begin{figure}
\centering
\includegraphics[width=.9\columnwidth]{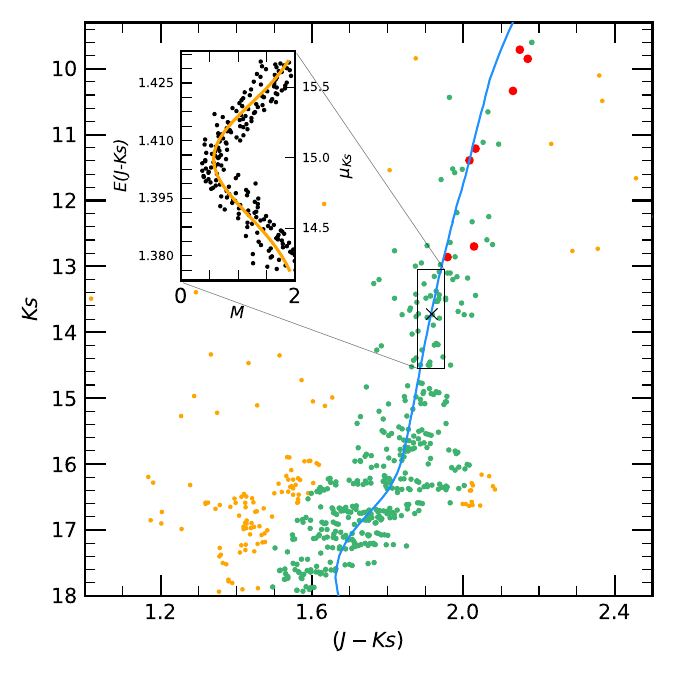}
\caption{CMD of VVV-CL001. Green points represent stars retained after the re-decontamination process, while orange points indicate those rejected. Red points correspond to CaT-confirmed members of VVV-CL001. The black cross marks the adopted fiducial point, and the blue curve shows the best-fitting theoretical isochrone. {\it Inner panel}: Black points show the value of the metric $M$ (Eq. \ref{eq:weight}) for each tested combination of $E(J-Ks)$ and $\mu_{Ks}$. The orange curve is the Gaussian fit to these points, used to identify the optimal values of $E(J-Ks)$ and $\mu_{Ks}$ that minimize $M$.}
\label{f_iso-fit}
\end{figure}

The age of VVV-CL001 was determined through isochrone fitting to near-infrared CMDs constructed using $J$, $H$, and $Ks$ photometry (Section \ref{ss_cmd}), in combination with theoretical isochrones from the \textit{PARSEC} database \citep{bressan_2012}. A full description of the method is provided in Appendix \S\ref{appendix_A}; we summarize the key aspects here.

Isochrone fitting depends on four parameters: age, metallicity, apparent distance modulus ($\mu_{Ks}$), and reddening ($E(J-Ks)$). In our analysis, the metallicity was fixed at $[\mathrm{Fe/H}] = -2.25 \pm 0.01$~dex, as derived in Section \ref{ss_met}. To reduce degeneracies, $E(J-Ks)$ and $\mu_{Ks}$ were parametrized relative to a fiducial point (FP) placed on the RGB, where the sequence is well-defined and age-insensitive. The adopted FP coordinates are $(J-Ks, Ks) = (1.92, 13.72)$ mag (Fig. \ref{f_iso-fit}, black cross).

We employed an iterative \textit{convergent approach} to determine the best-fitting isochrone. Initial grids of theoretical isochrones spanning 7--13 Gyr were shifted across a range of $E(J-Ks)$ and $\mu_{Ks}$ values consistent with the FP. The optimal combination of $E(J-Ks)$ and $\mu_{Ks}$ was identified by minimizing the metric $M$ (Eq.~\ref{eq:weight}), which quantifies the distance between the shifted isochrone and the observed CMD (inner panel of Fig.~\ref{f_iso-fit}). With these values fixed, the best-fitting age was determined by minimizing the absolute mean residual $|\bar{R}|$ (Eq.~\ref{eq:r}), computed along the extinction vector and restricted to the SGB ($J-Ks < 1.8$ mag) to maximize age sensitivity. The process was repeated iteratively until the age varied by less than 0.1 Gyr between iterations, and consistent results were obtained whether starting from 7 Gyr or 13 Gyr — hence the term \textit{convergent approach}.

\begin{figure}
\centering
\includegraphics[width=\columnwidth]{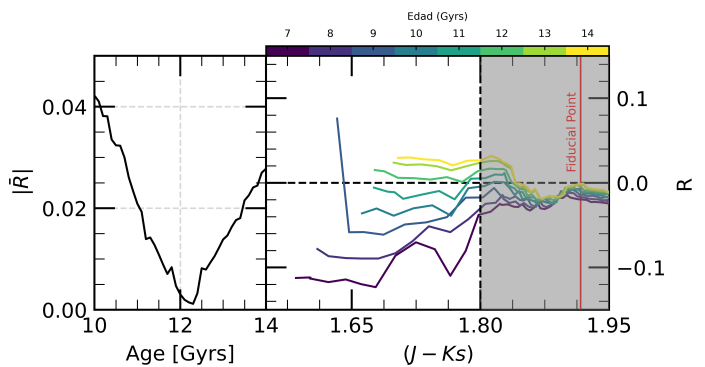}
\caption{{\it Left panel}: mean absolute residual $|\bar{R}|$ as a function of age for isochrones from 7 to 14 Gyr in steps of 0.1 Gyr. The minimum at 12.1 Gyr indicates the best-fitting age. {\it Right panel}: residual $R$ as a function of $J-Ks$ color for isochrones in 1 Gyr steps. The shaded region ($J-Ks > 1.8$ mag) was excluded from the age fit.}
\label{f_iso-fit2}
\end{figure}

The final best-fitting isochrone is overplotted on the CMD in Fig. \ref{f_iso-fit}. We derive an age of $12.1^{+1.0}_{-1.2}$ Gyr, with reddening $E(J-K_s)=1.40^{+0.01}_{-0.02}$ mag and distance modulus $\mu_{Ks}=15.01^{+0.19}_{-0.48}$ mag. The left panel of Fig. \ref{f_iso-fit2} shows the behavior of $|\bar{R}|$ as a function of age, with a clear minimum at 12.1 Gyr. The right panel illustrates the variation of the residual $R$ with color for a subset of isochrones; the shaded region ($J-Ks > 1.8$ mag) was excluded from the fit due to the weaker age sensitivity of the RGB.

Uncertainties in the age were estimated by repeating the fitting process while varying the $E(J-Ks)$, $\mu_{Ks}$, and metallicity within their respective errors. The individual contributions to the age uncertainty were determined as follows: metallicity variations yielded an uncertainty of $^{+0.3}_{-0.2}$ Gyr, while the combined variation of $E(J-Ks)$ and $\mu_{Ks}$  which are correlated parameters  resulted in an uncertainty of $^{+1.04}_{-1.22}$ Gyr. The final age uncertainty was computed as the quadratic sum of these deviations, giving $^{+1.0}_{-1.2}$ Gyr.

Although a reddening gradient toward the Galactic center was identified in Section \ref{ss_cmd}, tests confirmed that its effect on our results is negligible and already accounted for in the error estimates. We note that our photometry does not reach the main-sequence turnoff, but does include the SGB, which provides a more reliable age diagnostic  than previous results. Our final age is consistent within 1$\sigma$ with the value of $11.9^{+3.12}_{-4.05}$ Gyr reported by \citetalias{Fernandez21}.

\subsection{Distance}
\label{ss_distance}

The cluster distance can be derived from the apparent distance modulus $\mu_{Ks}$ and the reddening $E(J-Ks)$ only if the extinction law along the line of sight is well constrained \citep{Cardelli_1989}. However,
fluctuations of the density and composition of interstellar dust in the direction of the Galactic bulge \citep{nataf_2010} introduce significant variations of the extinction law, and they complicate photometric measurements and precise distance estimates \citep{Minniti_2010, Gonzalez_2012}. Although NIR extinction is lower than in optical wavelengths, it remains a crucial factor that must be carefully modeled to minimize its effects on photometry and distance determination \citep{Indebetouw_2005}.

\begin{figure}
\centering
\includegraphics[width=\columnwidth]{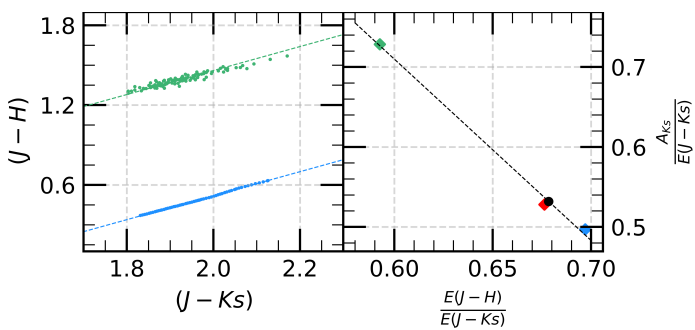}
	\caption{{\it Left Panel}: color-color diagram. Green points: only RGB stars from the photometric data resulting from statistical decontamination. Blue points: the best-fit isochrone model of Section~\ref{ssec:age}. The green and blue dotted curves show the linear fit of the green and blue points, respectively.  {\it Right Panel}: plane of $\frac{E(J-H)}{E(J-Ks)}$ vs. $\frac{A_{Ks}}{E(J-Ks)}$. Green diamond: standard extinction law of \cite{Cardelli_1989} converted to the 2MASS system according to Table 2 of \cite{catelan_2011}. Red and blue diamonds: bulge extinction law of the N$^{b}_{+}$ quadrant of \cite{Nishiyama_2009} and the average of their four quadrants, respectively. The dotted curve shows the linear fit of these three values. Black point: interpolated value for VVV-CL001.}
\label{f_ccd}
\end{figure}

The left panel of Fig. \ref{f_ccd} shows our determination of the color excess $E(J-H)$ by comparing the observed cluster RGB with the theoretical isochrone in the $(J-H)$ vs. $(J-K_s)$ diagram. The theoretical isochrone corresponds to the best-fitting model from Section \ref{ssec:age} and includes evolutionary stage labels (1: Main Sequence (MS), 2: SGB, 3: RGB, 4: Asymptotic Giant Branch). We corrected the isochrone for the adopted values of $E(J-Ks)$ and distance modulus $\mu_{Ks}$ determined in Section \ref{ssec:age}, which allowed us to estimate the RGB sequence in the observational data. The vertical separation between the reddening-corrected observational data and this isochrone returned $E(J-H)=0.95\pm0.01$. This analysis confirms the nearly linear trend of both RGB and SGB sequences in the color-color diagram, though we note that potential spurious detections in the SGB region could slightly affect its linear fit.

Since VVV-CL001 is located in the Galactic bulge, the standard extinction law of \cite{Cardelli_1989} is not suitable. Instead, the extinction law from \cite{Nishiyama_2009} is more appropriate. However, this study does not cover the specific region in which VVV-CL001 is located. To determine the most applicable extinction law for the cluster, we interpolated the values of $A_{K_s}/E(J-K_s)$ given in these two studies, as a function of $E(J-H)/E(J-K_s)$, as shown in Fig.~\ref{f_ccd}. This interpolation yields $A_{K_s}/E(J-K_s)=0.53\pm0.01$ for the value $E(J-H)/E(J-K_s)=0.68\pm0.01$ previously obtained for VVV-CL001. Our result is consistent with the average of the four quadrants studied by \cite{Nishiyama_2009}, but it deviates from the quadrant that is closest to the position of the cluster.

Finally, the cluster's distance was determined using the distance modulus relation:
\begin{equation}
	\mathrm{d}_\odot = 10^{[\mu_{Ks}+5-E(J-K_s)\frac{A_{Ks}}{E(J-K_s)}]/5},
\end{equation}
with $\mu_{Ks}$ and $E(J-K_s)$ obtained in Section~\ref{ssec:age}, $A_{K_s}/E(J-K_s)$ determined in this section, $A_{Ks}$ the extinction in the $K_s$ band, and $\mathrm{d}_\odot$ the distance in parsecs. The cluster distance was computed as $\mathrm{d}_\odot = 7.1^{+1.3}_{-1.1}$ kpc. The uncertainty was estimated following a similar procedure as in Section \ref{ssec:age}. The individual contributions to the distance uncertainty were determined as follows: metallicity variations yielded an uncertainty of $\pm 0.2$ kpc, while the combined variation of $E(J-Ks)$ and $\mu_{Ks}$ which are correlated parameters resulted in an uncertainty of $^{+1.3}_{-1.1}$ kpc. The final distance uncertainty was computed as the quadratic sum of these deviations. Our distance estimate is in agreement within the error bars with that reported by \citetalias{Fernandez21}, who determined a distance of $8.22^{+1.84}_{-1.93}$ kpc. Similarly, we achieve agreement within the error bars with what was reported by \citetalias{Olivares22}, who determined a distance of $8.23\pm0.46$ kpc. However, it should be noted that the latter mentions that their determination should be considered an upper limit for the distance of VVV-CL001. The discrepancies between the mean values we found and those in the literature may be primarily due to the methodology used in these studies, as well as the extinction law applied in each estimation. The latter, in particular, can cause significant differences with small variations in the extinction value used.

\begin{table}[t]
\begin{center}
\renewcommand{\arraystretch}{1.2} 
\caption{Derived cluster parameters.}
\label{t_CL001res}
\begin{tabular}{l c l}
\hline
\hline
(RA, dec) & ($17^h 54^m 42\fs22$, $-24\degr 00\arcmin 52\farcs3$) & \\
$r_t$ & $3\farcm48$ & \\
c & 0.57 & \\
$(\mu_\alpha^*,\mu_\delta)$ & $(-3.68\pm0.09, -1.76\pm0.10)$ & mas~yr$^{-1}$ \\
RV & $-334\pm4$ & km~s$^{-1}$ \\
$[\mathrm{{Fe}/{H}}]$ & $-2.25\pm0.05$ & dex \\
E($J-K_\mathrm{s}$) & $ 1.40^{+0.01}_{-0.02}$ & mag \\
d$_\odot$ & $7.1^{+1.3}_{-1.1}$ & kpc \\
r$_{\rm GC}$ & $1.3^{+0.9}_{-0.5}$ & kpc \\
$Z_{\rm GC}$ & $102^{+13}_{-12}$ & pc \\ 
Age & $12.1^{+1.0}_{-1.2}$ & Gyr \\
\hline
\end{tabular}
\end{center}
\end{table}

\section{Orbital properties}
\label{s_orbit}

We derived the main orbital properties of VVV-CL001 using the observational estimates in Table \ref{t_CL001res} as initial conditions, and for the orbital calculations we used the software \textsc{delorean} \citep{Blana2020}. 
This code has an updated MW potential (Bla\~na et al. in prep.) that includes the MW bulge/bar stellar mass component of \citet[][hereafter S22]{Sormani2022} \defcitealias{Sormani2022}{S22}, which was derived from made-to-measure models fitted to observations \citep{Wegg2013,Portail2017a}.
This bar model consists of an inner disky component, a thick box/peanut bulge structure extending $\sim$2 kpc in radius and $\sim$1 kpc in height, a flat/thin bar component that extends to 5 kpc in radius and 45 pc height, and another stellar disk \citep[][hereafter BG16]{Bland-Hawthorn2016} \defcitealias{Bland-Hawthorn2016}{BG16}.
We adopted the bar pattern speed angular frequency of $\mathbf{\Omega}_{\rm bar} = (-) 39.0\pm3.5 \, \hat{Z} \,{\rm km\,s^{-1}\,kpc^{-1}\, rad}$ (equivalent to a rotation period of $T_{\rm bar}\!=\!157\,{\rm Myr}$) \citep{Portail2017a} and a bar angle of $-27^{\rm o}$ \citep{Wegg2013} which encompasses other estimates in the literature within errors. The potential model also includes components for the dark matter halo, gaseous disk, and nuclear components. In Appendix \S\ref{sec:app:orb}, we provide additional information, as well as the circular velocity profiles of the potential model (see Fig.\ref{fig:MWvc}). For the solar parameters, we adopted the values of \citetalias{Bland-Hawthorn2016} for the Sun's galactocentric distance of $R_{\rm 0}\!=\!8.2\pm0.1\,{\rm kpc}$, a vertical height of $z_{\rm 0}\!=\!25\pm5\,{\rm pc}$, and the solar constants of motion of $(u_{\odot}, v_{\odot}, w_{\odot})=(1\!\pm\!1,\, 248\!\pm\!3, \,7.3\!\pm\!0.5)\,{\rm km/s}$. With this we estimate that the current galactocentric distance of VVV-CL001 is $r_{\rm GC}\!=\!1.3^{+0.9}_{-0.5}\,{\rm kpc}$ and its distance from the Galactic mid-plane is $Z_{\rm GC}\!=\!101^{+13}_{-12}\,{\rm pc}$, with errors derived from a Monte Carlo propagation of the errors of the cluster coordinates and the solar constants.

\begin{figure}[ht]
\centering
\includegraphics[width=.9\columnwidth]{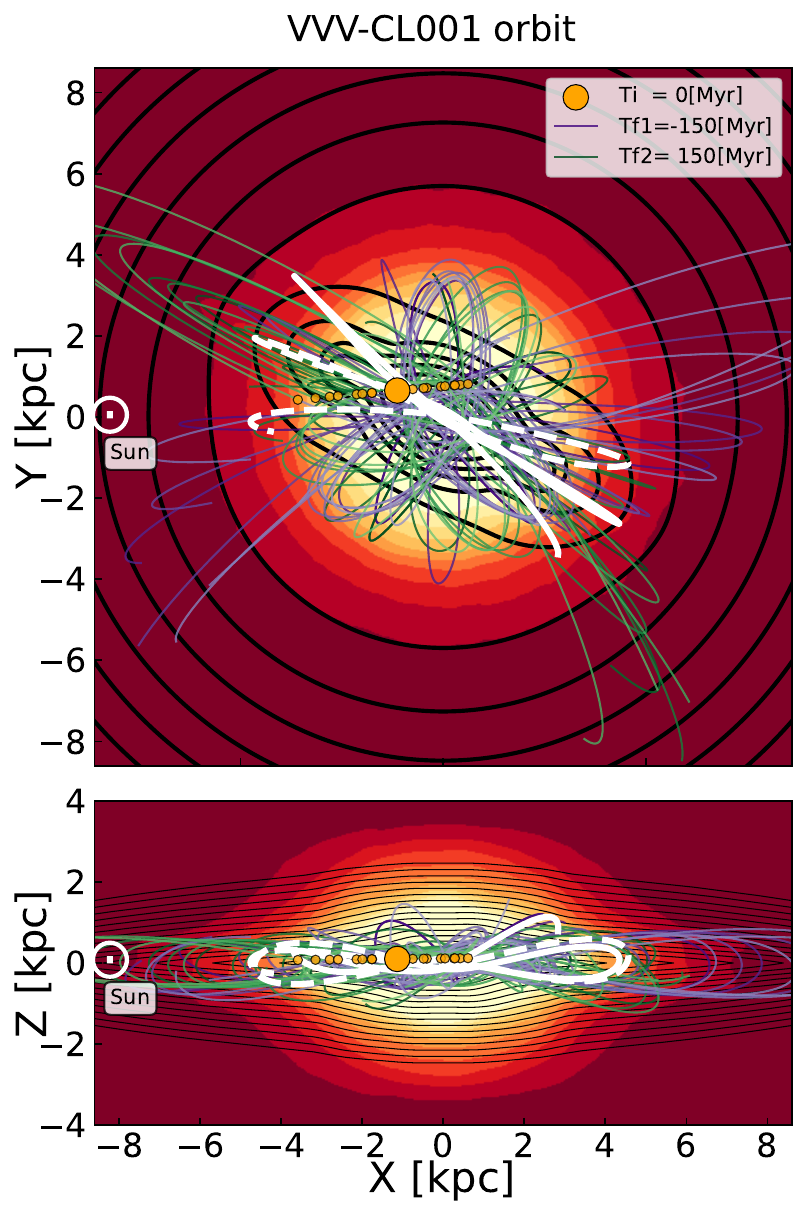}
\caption{
Orbital calculations for VVV-CL001. 
We show the most likely current position (large orange circle) and the corresponding orbit (white curve) showing its past -150 Myr orbit (solid) and future 150 Myr orbit (dashed) seen in the $X\!-\!Y$ plane view (top panel) and the $X\!-\!Z$ view (bottom panel).
To avoid crowding we plot a sub-sample of 20 randomly selected orbits from the total of $10^4$ orbits, showing their current positions (smaller circles), their past orbits (green) and future orbits (purple).
The density probability (color map) shows the regions that the orbits inhabit.
The stellar surface mass density contours of the MW bar and inner disk model of \citetalias{Sormani2022} reveal that this cluster lives between the inner disk and the bar/bulge region.
We also plot the Sun's position. We note here that in this frame the MW and the bar rotate clock-wise, with the bar having a rotation period of $T_{\rm bar}\!=\!157\,{\rm Myr}$.
}
\label{fig:orbits}
\end{figure}

Considering this set-up for the potential and the cluster coordinates, we computed the orbital properties of VVV-CL001. To explore the effects of uncertainties, we adopted a Monte Carlo method by randomly sampling $10^4$ values from normal distributions with standard deviations with sizes taken from the estimated errors in the coordinates shown in Table \ref{t_CL001res}. 
We also sample values for the bar pattern speed to account for this uncertainty.
We took the orbital integration time of 6\,{\rm Gyr} ($5\,{\rm Gyr}$ into the past and $1\,{\rm Gyr}$ into the future). While arbitrary, this time is long enough to consider multiple orbital timescales for the cluster to explore phase-space, where we find typical orbital periods of $T_{\phi}\!\sim\!100\,{\rm Myr}$. 
Longer orbital integrations would be comparable to changes in the MW potential evolution and/or possible orbital interactions, which would propagate these time-dependent uncertainties into the orbital parameters. 
Furthermore, we tested longer or shorter integration time intervals, finding similar results.
From the resulting orbital distributions, we calculated the medians and 1-sigma confidence values of the orbital properties such as the pericenter ($r_{\rm peri}$), the apocenter ($r_{\rm apo}$), the maximum vertical height ($|Z|_{\rm max}$), eccentricity ($e=(r_{\rm apo}-r_{\rm peri})(r_{\rm apo}+r_{\rm peri})^{-1}$), and additional relevant orbital properties. 
These properties are then compared with the current structural properties of the MW bulge, bar and disk.
We note that in this reference system, where the Sun is located at $X\!=\!-8.2\,{\rm kpc}$, the MW disk and bar have a clockwise rotation and, therefore, they have negative $Z$-axis specific angular momentum ($lz<0\,{\rm kpc\,km/s}$).

\begin{table}[t]
\begin{center}
\caption{VVV-CL001 orbital parameters.}
\vspace{-0.2cm}
\normalsize
\renewcommand{\arraystretch}{1.6} 
\setlength{\tabcolsep}{4pt}      
\begin{tabular}{c c c c c c c c}
\hline\hline
$r_{\rm apo}$ & $r_{\rm peri}$ &  $e$ & $|Z|_{\rm max}$ & $T_{R}$ & $T_{\phi}$ & $T_{Z}$ \\
${\rm [kpc]}$         &    ${\rm [kpc]}$            & -       & ${\rm [kpc]}$           & ${\rm [Myr]}$  & ${\rm [Myr]}$       & ${\rm [Myr]}$\\
\hline
$4.5^{+2.5}_{-1.2}$  & $0.6^{+0.3}_{-0.2}$  &   $0.76^{+0.10}_{-0.14}$ & $1.03^{+0.45}_{-0.32}$  & $56^{+20}_{-12}$  & $105^{+45}_{-32}$  & $71^{+20}_{-12}$ \\
\hline
\end{tabular}
\label{tab:orbparams}
\begin{tablenotes}
Note: medians and 1-$\sigma$ percentile intervals of orbital parameters, identified from left to right: (1) apocenter, (2) pericenter, (3) eccentricity, (4) the maximum height above (below) the MW mid-plane, (5) cylindrical radial orbital period, (6) azimuthal period, (7) vertical period.
\end{tablenotes}
\end{center}
\end{table}

We present the orbital computations of VVV-CL001 in Fig. \ref{fig:orbits}, showing the orbit with the most likely position and a subsample of randomly selected orbits. 
We used the total sample of $10^4$ orbits to calculate the distribution of the orbital parameters, presenting   their medians and 1-$\sigma$ percentile intervals in Table \ref{tab:orbparams}.
An inspection of Fig. \ref{fig:orbits} of the sample of possible orbits and the density probability maps
reveals that the cluster lives orbiting the volume occupied by the bar and inner disk.
It enters the disk region when it approaches its orbital apocenter ($r_{\rm apo}\!=\!4.5\,{\rm kpc}$), and it enters the bar and central bulge when it reaches its pericenter ($r_{\rm peri}\!=\!0.6\,{\rm kpc}$). Moreover, we find that the orbits remain vertically close to the disk and box/peanut bulge of the MW, extending only out to $|Z|_{\rm max}=1\,{\rm kpc}$. It orbits around the MW every $T_{\phi}\!=\!107\,{\rm Myr}$, while it enters the central region of the box/peanut bulge every $T_{R}\!=\!55\,{\rm Myr}$, approaching the epicycle approximation ($T_{\phi}\sim 2T_{R}$), and a vertical period around the midplane every $T_{Z}\!=\!71\,{\rm Myr}$.

We also measured the specific angular momentum of the orbit of the cluster around the MW's rotation axis ($lz$) finding that it is not conserved, which is expected for a triaxial rotating potential such as the MW bar. 
We measured whether the cluster orbit is prograde ($lz<0\,{\rm kpc\,km\,s^{-1}}$) or retrograde ($lz>0\,{\rm km\,s^{-1}\,kpc}$) with respect to the MW (here $lz_{MW} < 0\,{\rm km\,s^{-1}\,kpc}$). 
We find that with the current observed positions and velocities distributions $lz$ has a median and deviations of $lz_{\rm obs}=105^{+155}_{-186}\,{\rm km\,s^{-1}\,kpc}$, finding that within the observational errors, we find pro-grade (28\%) and retro-grade (72\%) orbital directions.
However, given that the cluster constantly crosses between the outer and inner regions of the bar near the mid plane, the rotating potential can perturb the orbits generating a chaotic behavior that can frequently flip the direction of rotation of the orbit, making prograde-retrograde orbits. 
Therefore, the median of $lz$ covers prograde and retrograde phases with $lz_{\rm med}\!=\!28^{+148}_{-150}\,{\rm km\,s^{-1}\,kpc}$, with minima of $lz$ having a range of $lz_{\rm min}\!=\!-39^{+147}_{-186}\,{\rm km\,s^{-1}\,kpc}$, while the maxima have $lz_{\rm max}\!=\!128\pm150\,{\rm km\,s^{-1}\,kpc}$.
This chaotic orbital behavior has already been identified in other star clusters in the MW \citep[e.g.][]{Pichardo2004,Perez-Villegas2018,Perez-Villegas2019,Tkachenko2023}.

We also performed tests computing the orbital properties in different potentials, using an adaptation from \cite{plotnikova2023very}, finding similar values for the apocenter ($r_{\rm apo}\!=\!6.4\pm1.3\,{\rm kpc}$), the pericenter ($r_{\rm peri}\!=\!0.37\pm0.16\,{\rm kpc}$), and a vertical height of $|Z|_{\rm max}\!=\!0.65\pm0.09\,{\rm kpc}$. The differences arise due to differences of the bar mass model.
To test the effects of the bar on the orbital properties, we implemented the axisymmetric MW potential MWPotential2014 from \citet{Bovy2015} into \textsc{delorean}. 
We find similar overall orbital properties, such as the orbital periods, with some differences due to conserved quantities (e.g. $lz$) allowed by the barless potential, producing orbits more confined to the plane of the galaxy with $|Z|_{\rm max}=430^{+90}_{-70}\,{\rm pc}$, reaching pericenter and apocenter means of $r_{\rm per}=0.41^{+0.46}_{-0.17}\,{\rm kpc}$ and $r_{\rm apo}=5.1^{+2.3}_{1.1}\,{\rm kpc}$, respectively.

Consequently, considering the orbital distribution of the cluster that concentrates its orbit around the box/peanut bulge and bar within $r\!\sim\!5\,{\rm kpc}$ while reaching the central region around $r\!\sim\!0.6\,{\rm kpc}$, its relative confinement on the vertical axis ($|Z|_{\rm max}\!\sim\! 1\,{\rm kpc}$), its old age ($12.1\,{\rm Gyr}$) and metal-poor content, we propose a scenario where this cluster was formed in the very early disk before the formation of the bar. 
Then it was captured during the formation of the bar, between 8 and 10 Gyr ago \citep[][see also \citet{Schoedel2023}]{Sanders2024}, and it got pulled into higher orbits in $Z$ through heating mechanisms, such as bar buckling and satellite interactions.
We also compared the orbital properties of VVV-CL01 with the GC orbital classification of \citet[][see their Fig.6]{Perez-Villegas2019}, finding that this cluster would agree with both thick disk and the bulge/bar GC populations.

We also compared with specific studies of VVV-CL001 in the literature. \citetalias{Fernandez21} also explored a vast parameter space, presenting different orbital solutions that depend on the adopted observational coordinates of the cluster, finding some prograde-retrograde solutions, and concluding that the star cluster could be associated with halo-like orbits and accretion events. In general, we find good agreement with their orbital calculations, particularly with their model with a cluster distance of $d_{\odot}=7.87-8.10\,{\rm kpc}$, which results in a Jacobi (energy) integral and orbital energy values that can also be associated with the box/peanut bulge (see their Fig.4). On the other hand, the model of \citet{Vasiliev21} produced more compact orbits of $r_{\rm apo}\!\sim\!2\,{\rm kpc}$, always retrograde and regular orbits, which is due to the compact (barless) spherical bulge potential used in their models \citep[][MWPotential2014]{McMillan2017, Bovy2015}, and due to differences in the observed coordinate values. Using the same models for the potential, \citetalias{Olivares22} determined that for this cluster, $r_{\rm apo}\!\sim\!3.2\,{\rm kpc}$, and given its position and very low velocity, it suggests that it is associated with the bulge or inner halo rather than with the thick disk. Moreover, in addition to the MW potential, our error propagation study shows that the cluster's distance is the most critical variable that determines its orbital history, as this cluster is so close to the bulge center ($r_{\rm GC}\!=\!1.3^{+0.9}_{-0.5}\,{\rm kpc}$) where the potential is deepest, deviations of $\pm1\,{\rm kpc}$ significantly change its orbital energy, as well as the conversion from proper motions to physical velocities.

\section{Discussion and Conclusions}
\label{Dis_concl}

Our study provides a detailed characterization of the unique GC VVV-CL001, establishing it as an ancient \textit{in-situ} and metal-poor object located in the Galactic bulge region.

Our results show both agreements and discrepancies with previous studies. The derived coordinates differ by $1\farcs8$ compared to \cite{Minniti11}. The structural parameters obtained, with a core radius of $r_c = 0\farcm94$ and a tidal radius of $r_t = 3\farcm48$, suggest that it is a relatively compact cluster.

Regarding kinematics, the proper motions obtained are consistent within $1\sigma$ with those of \cite{Vasiliev21} and \citetalias{Fernandez21}, but show an irreconcilable difference in the component of $\mu_\delta$ with those of \citetalias{Olivares22} (see Table~\ref{t_PM}). The radial velocity obtained in this work, $RV = -334 \pm 4~\mathrm{km\,s^{-1}}$, is similar to the values reported by \citetalias{Fernandez21} and \citetalias{Olivares22}, who found $\sim -325~\mathrm{km\,s^{-1}}$ and $-324.9 \pm 0.8~\mathrm{km\,s^{-1}}$, respectively. Methodological differences between these works could explain the small discrepancies. The metallicity value of $\mathrm{[Fe/H]} = -2.25 \pm 0.05$ obtained in this study is consistent with that of \citetalias{Fernandez21} but not with that of \citetalias{Olivares22}, who report $-2.45 \pm 0.24$ and $-2.04 \pm 0.02$, respectively.

However, we find significant differences in the reddening affecting the cluster. Our value of $E(J-K_S) = 1.40$ differs from $\sim 1.34$ adopted by \citetalias{Olivares22} in the determination of their distance. The distance derived in this work, $d_\odot = 7.1$ kpc, agrees within $1\sigma$ with that of \citetalias{Fernandez21} but not with that of \citetalias{Olivares22}. However, \citetalias{Olivares22} note that their reported distance should be considered an upper limit for VVV-CL001, and under this context, our results are consistent. Finally, our age estimate of $12.1^{+1.0}_{-1.2}$ Gyr is consistent with that of \citetalias{Fernandez21}, while \citetalias{Olivares22} did not report an age estimate.

Based on our results,we can explore the possible origin of VVV-CL001. Its low metallicity and age suggest that it formed at a time when the interstellar medium had not yet been significantly enriched by supernova Ia explosions \citep{marin2009acs, Lahen_2019,chiti_2021}. However, to confirm this hypothesis, a more detailed spectroscopic analysis of the cluster's stellar population is required. In this sense, conducting high-resolution near-infrared spectroscopic studies, similar to that of \citetalias{Fernandez21} but for a larger sample of stars, would be crucial to continue investigating the early formation and evolution processes of this GC and, consequently, help to recover the formation history of the MW.

We also calculated the orbital properties of VVV-CL001 with updated Galactic potentials (see Section \ref{s_orbit}). We find that the cluster has a prograde-retrograde type orbit due to the barred potential \citep[see ][]{tkachenko23}. The orbits are eccentric ($e\!=\! 0.76$), traveling between the box/peanut bulge central region ($\mathrm{r_{peri}} \!=\! 0.6$ pc) and the flat bar and disk region ($\mathrm{r_{apo}} \!=\! 4.5$ kpc). Moreover, the orbit lies confined within $|Z|_{\rm max} \sim 1$ kpc from the disk mid-plane, suggesting a connection with this structure.

These orbital properties strongly suggest an \textit{in-situ} origin in the early proto-Galactic disk, before the formation of the bar. It is likely that the cluster was captured by the bar's gravitational potential in its early stages, being vertically lifted from the galactic midplane through dynamical mechanisms, such as the bar buckling that formed the boxy/peanut bulge, or satellite interactions. Thus, VVV-CL001 would be a fossil remnant from the earliest phases of Galactic assembly and a valuable candidate for being part of the population that gave rise to the inner disk and bulge. It is currently the most metal-poor in-situ GC known with a well-determined metallicity \citep{geisler25}, and further study will help reveal conditions in the early history of the main progenitor of the MW.

The main uncertainties in this study arise from the lack of our knowledge of the extinction law for this region of the Galaxy, affecting the precise distance determination. Additionally, the photometry used was limited by spurious detections in the catalog and the inability to accurately characterize the cluster's turnoff, which limits the precise age estimation.

To improve the characterization of VVV-CL001, deeper photometry would be essential, allowing a more accurate determination of the cluster's age and other evolutionary parameters. Moreover, the development of a more detailed extinction law for the Galactic disk and bulge region would be crucial to reduce uncertainties in the distance determination, as this is the parameter that most strongly affects age estimation according to \cite{ying2023}), given sufficiently deep photometry well below the MSTO. We note that we are currently obtaining very deep J,$\mathrm{K_s}$ images for VVVCL1 from the GEMINI-S GeMS + GSAOI instrument, which should reach several magnitudes below the MSTO and provide a much more precise and accurate age.

This work represents one of the most comprehensive studies to date on VVV-CL001, combining astrometric, photometric, and spectroscopic information to characterize its nature. The methodology used to determine its age, distance, and metallicity provides a solid framework for future studies of GCs in highly extincted regions of the MW. Finally, the possibility that VVV-CL001 is a relic from the early stages of Galactic formation makes it an ideal candidate for more detailed studies, combining proper motion information with high-resolution spectroscopic data. It may well be the oldest remnant of the MW main progenitor, and a reliable age and more detailed chemistry will help constrain the formation and early chemical evolution of the main progenitor and place a lower limit to the age of the MW and indeed the Universe.

\begin{acknowledgements}
    This work has made use of data from the European Space Agency (ESA) mission {\it Gaia} (\url{https://www.cosmos.esa.int/gaia}), processed by the {\it Gaia} Data Processing and Analysis Consortium (DPAC, \url{https://www.cosmos.esa.int/web/gaia/dpac/consortium}). Funding for the DPAC has been provided by national institutions, in particular the institutions participating in the {\it Gaia} Multilateral Agreement. 
    D.G. gratefully acknowledges the support provided by Fondecyt regular no. 1220264.
    D.G. also acknowledges financial support from the Direcci\'on de Investigaci\'on y Desarrollo de la Universidad de La Serena through the Programa de Incentivo a la Investigaci\'on de Acad\'emicos (PIA-DIDULS).
    C.M. thanks the support provided by ANID-GEMINI Postdoctorado No.32230017.
    S.V. gratefully acknowledges the support provided by Fondecyt Regular n. 1220264 and by the ANID BASAL project FB210003.
    B.D. acknowledges support from ANID Basal project FB210003.
    M.B. thanks M.A. Perez Villegas and M. De Leo for helpful discussions.
\end{acknowledgements}

\bibliographystyle{aa}
\bibliography{bibtex/CL001bib}

\begin{appendix}
\nolinenumbers

\section{Isochrone fit: considerations and method}
\label{appendix_A}

This section describes in detail the isochrone fitting method used in this work. The procedure is based on minimizing the distance between the points in the CMD and the theoretical isochrones. A convergent approach was adopted to identify the optimal combination of parameters that best describe the stellar cluster. In this context, the method is considered converged when, starting from different sets of initial parameters, the final results agree within the error margins.

An isochrone fitting must consider four fundamental parameters: age, metallicity, distance, and reddening. Age and metallicity are intrinsic parameters of the cluster, directly related to the morphology of the isochrone. In contrast, distance and reddening are extrinsic parameters that depend on the cluster’s position relative to the observer, affecting the photometric distribution of stars in the CMD. For this reason, it is essential to determine the extrinsic parameters first, and then isolate and accurately estimate the cluster’s intrinsic parameters.

In this particular study, one of the intrinsic parameters metallicity was previously determined in Section \ref{ss_met}. Therefore, the fitting was performed by adopting a fixed metallicity and exploring a wide range of ages, from 7 to 14 Gyr. The following subsections detail the considerations and steps adopted in the isochrone fitting method.

\subsection{Consideration: theoretical isochrones}

Theoretical isochrones were built using version 1.2S of the stellar evolution code \texttt{PARSEC} \citep{bressan_2012}. The metallicity used for generating the isochrones was computed using the equation proposed by \citet{Salaris_1993}, which relates the global metallicity [M/H] to the metallicity [Fe/H] and the alpha-element abundance [$\alpha$/Fe]:

\begin{equation}
    [M/H] = [Fe/H] + \log(0.638 \times 10^{[\alpha/Fe]} + 0.362)
\end{equation}

In this work, we adopted $[Fe/H] = -2.25 \pm 0.05$, determined in Section \ref{ss_met}, and [$\alpha$/Fe] = 0.335 dex, obtained from the mean abundances of O, Mg, and Si reported by \citetalias{Fernandez21} for two cluster member stars. With these values, the global metallicity is $[M/H] = -1.98 \pm 0.12$ dex. The ages considered for constructing the theoretical isochrones range from 7 to 14 Gyr, in steps of 0.1 Gyr, covering a representative range of typical GC ages.

\begin{figure}
\centering
\includegraphics[width=.9\columnwidth]{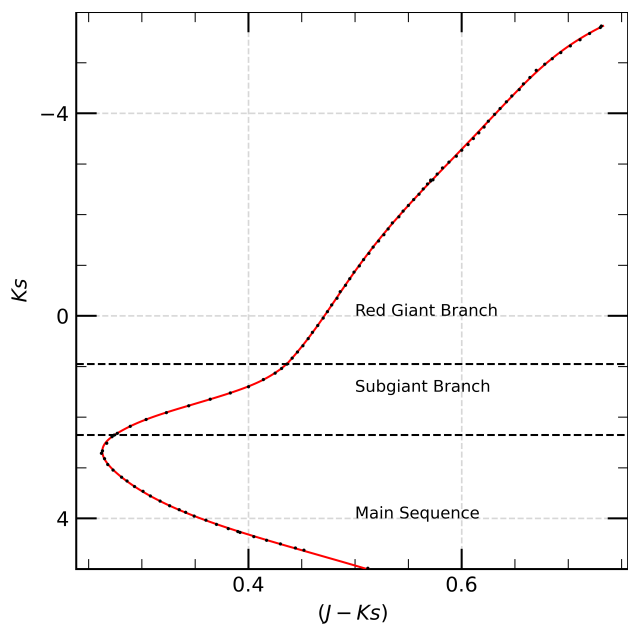}
	\caption{ Example of the polynomial fitting applied to a theoretical isochrone. The black points correspond to the discrete output of the \texttt{PARSEC} code, while the red line represents the polynomial fit used to generate a uniform distribution of points along the sequence. The isochrone shown corresponds to an age of 12.1~Gyr and to the metallicity of the cluster.
    }
\label{fig:iso_poly}
\end{figure}

It is important to note that the theoretical isochrones extracted from \textit{PARSEC} are not continuous curves, but discrete and non-uniform sets of points (see Fig. \ref{fig:iso_poly}). This feature may introduce biases if directly compared with observational data. To mitigate this effect, a polynomial was fitted to each of the principal sequences of the theoretical isochrone MS, SGB, and RGB to generate a uniform array of points and avoid distortions caused by the non-uniformity of the original models.

Legendre polynomials were used for this fitting. In addition, the difference in magnitude between consecutive points was restricted to a maximum of 0.02 mag. Although this restriction implies a slight non-uniformity in color, the effect is minimal: in the RGB region used to determine the reddening (see Section \ref{ss:fp}), the maximum color difference between consecutive points is only 0.0019 mag. The impact of this variation on the final reddening values is negligible, as it lies well within the associated error margins. Furthermore, its effect on the determination of age is less than 0.1 Gyrs, which is lower than the age range among the different models used.

Fig. \ref{fig:iso_poly} shows an example of the polynomial fitting applied to a theoretical isochrone. For practical purposes, in sections \ref{ss:fp}, \ref{ss:red&mu} and \ref{ss:redecon}, “theoretical isochrone” or simply “isochrone” will refer to the uniform array of points generated through this procedure, rather than the original discrete curve.

\subsection{Consideration: fiducial point (FP) and parameterization}
\label{ss:fp}

To facilitate the comparison between observational data and theoretical isochrones, a FP was defined in the CMD. This is based on the hypothesis that, when the theoretical model is properly fitted to the observed data, there must exist a point on the isochrone that coincides or closely matches the FP within photometric uncertainties. Although an exact correspondence is not always possible due to the discrete nature of both the models and the observations, the FP provides an appropriate initial reference for exploring the $E(J-Ks)$ and $\mu_{Ks}$ parameter space.

The FP was determined as the average magnitude and color in the CMD region defined by $13.6 < Ks < 13.8$ mag and $0.5 < J-Ks < 2.5$ mag, resulting in $(Ks, J-Ks) = (13.72, 0.92)$ mag. Fig. \ref{f_iso-fit} shows the position of the FP, marked with a cross.

Once the FP is defined, the difference in magnitude and color between this point and each point of the theoretical isochrone can be computed. These differences represent the $E(J-Ks)$ and $\mu_{Ks}$ values required to align the isochrone with the observational data, defined as:

\begin{equation}
\label{eq:correcciones}
\mu_{Ks,i} = Ks^{iso}_{i} - Ks_{FP}, \quad E(J-Ks)_{i} = (J-Ks)^{iso}_{i} - (J-Ks)_{FP},
\end{equation}

where $Ks^{iso}_{i}$ and $(J-Ks)^{iso}_{i}$ are the magnitude and color of the $i$-th point of the theoretical isochrone, while $Ks_{FP}$ and $(J-Ks)_{FP}$ correspond to the coordinates of the FP. This yields a set of values $\{\mu_{Ks,i}, E(J-Ks)_{i}\}$ representing the possible corrections to align the theoretical isochrone with the observations in the CMD. For a discussion of their effects on the determination, see Section \ref{ss:red&mu}.

\begin{figure}
\centering
\includegraphics[width=.9\columnwidth]{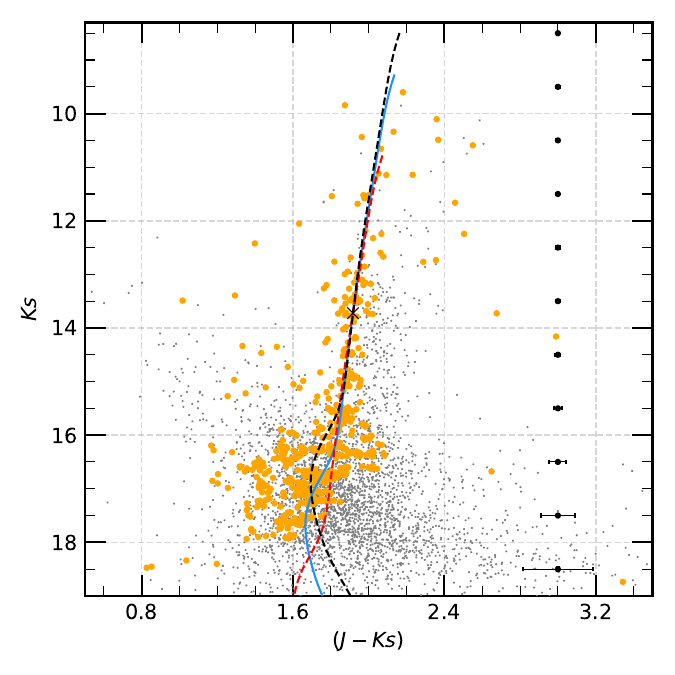}
	\caption{ Same as Fig. \ref{f_cmd} , but with overlaid isochrone curves to exemplify the fitting process. The solid blue curve represents the best-fitting isochrone resulting from the complete procedure. The dashed lines illustrate the same theoretical model shifted when choosing different comparison points; the red curve corresponds to a lower $E(J-Ks)_{i}$  value (1.36), while the black curve corresponds to a higher value (1.43) than that obtained for the optimal solution (blue curve).
    }
\label{fig:shift}
\end{figure}

Applying these corrections $\{\mu_{Ks,i}, E(J-Ks)_{i}\}$ can be geometrically interpreted as moving the points of the isochrone across the fiducial point. Figure \ref{fig:shift} exemplifies this concept

\subsection{\texorpdfstring{Reddening $E(J-Ks)$ and Distance modulus $\mu_{Ks}$}{Reddening E(J-Ks) and Distance modulus muKs}}

\label{ss:red&mu}
The adopted method follows a convergent approach to determine the best combination of parameters describing the stellar cluster. First, the extrinsic parameters $E(J-Ks)$ and $\mu_{Ks}$ are estimated, followed by the determination of the remaining intrinsic parameter, the age of the cluster, since the metallicity was already established in Section \ref{ss_met}.

The process begins by selecting an initial set of parameters, including the age of the theoretical isochrone and the previously defined fiducial point. As age is the last parameter to be determined, one of the extremes of the considered range (7 or 14 Gyr) is adopted as a starting point to test convergence toward the same final value regardless of the initial choice.

Then, the values $\mu_{Ks,i}$ and $E(J-Ks)_{i}$ are calculated for each point of the theoretical isochrone using Equation \ref{eq:correcciones}. Each pair $(\mu_{Ks,i}, E(J-Ks)_{i})$ is applied to shift the isochrone in the CMD (see Fig. \ref{fig:shift}), resulting in corrected points defined as:

\begin{equation}
\label{eq:iso_shifted}
Ks^{iso, corr}_{i} = Ks^{iso}_{i} - \mu_{Ks,i}, \quad (J-Ks)^{iso, corr}_{i} = (J-Ks)^{iso}_{i} - E(J-Ks)_{i},
\end{equation}

where $Ks^{iso, corr}_{i}$ and $(J-Ks)^{iso, corr}_{i}$ correspond to the magnitudes and colors of the shifted isochrone, respectively.  
For each shift, each observed star is associated with the nearest point of the shifted isochrone, creating groups of stars linked to the same theoretical point. Then, the mean and standard deviation of magnitude and color are computed for each group and compared with the isochrone values through a likelihood function defined as:

\begin{equation}
\label{eq:weight}
M = \frac{\sum_{j} \sqrt{ \left[ Ks_{j}^{iso}- \overline{Ks}_{j}^{obs} \right]^2 + \left[ (J-Ks)_{j}^{iso} - \overline{(J-Ks)}_{j}^{obs} \right]^2 }}{\sqrt{\sigma_{\overline{Ks}}^2 + \sigma_{\overline{(J-Ks)}}^2}},
\end{equation}

where $Ks_{j}^{iso}$ and $(J-Ks)_{j}^{iso}$ are the theoretical values of the $j$-th point of the shifted isochrone; $\overline{Ks}_{j}^{obs}$ and $\overline{(J-Ks)}_{j}^{obs}$ are the observational means of the associated set; and $\sigma_{\overline{Ks}}$ and $\sigma_{\overline{(J-Ks)}}$ are the standard deviations of the means. The sum extends over the isochrone points with associated observed stars.

This function naturally weights configurations that produce stellar sets with lower dispersion in magnitude and color, giving them greater importance. The goal of this step is to minimize $M$, which corresponds to maximizing the likelihood of the fit and thus identifying the optimal values $\mu_{Ks}^{opt}$ and $E(J-Ks)^{opt}$. The inner panel of Fig. \ref{f_iso-fit} illustrates the calculation and minimization of $M$ for a 12.1 Gyr isochrone.

An additional aspect to consider is the dependence of the real values of $\mu_{Ks}$ and $E(J-Ks)$ on the fiducial point. This can be assessed by examining the mean $\overline{(J-Ks)}_{j}^{obs}$ at the minimum of $M$, which indicates the average shift of the observational distribution relative to the isochrone. The final values of $\mu_{Ks}$ and $E(J-Ks)$ are thus obtained as:

\begin{equation}
\mu_{Ks} = \mu_{Ks}^{opt}, \quad E(J-Ks) = E(J-Ks)^{opt} + \delta_{E(J-Ks)},
\end{equation}

where $\delta_{E(J-Ks)}$ corresponds to the mean $\overline{(J-Ks)}_{j}^{obs}$ at the minimum of $M$. Therefore, the choice of the fiducial point is critical to avoid biases in determining the extrinsic parameters. Several points along the RGB were tested, yielding consistent values within the error margins. In this work, we adopted the previously defined FP, which produces $\delta_{E(J-Ks)}$ values on the order of 0.005 mag, thus minimizing any potential bias.

Finally, the obtained values of $\mu_{Ks}$ and $E(J-Ks)$ allow the calculation of the cluster distance and reddening as described in Section \ref{ss_distance}, providing the essential information to determine the extinction vector used in the cluster age estimation(See Section \ref{ss:AGE}).

\subsection{CMD re-decontamination}
\label{ss:redecon}

To remove stars that were not properly decontaminated in the initial process (see Section \ref{ss_cmd}), an additional CMD re-decontamination procedure was implemented. This method relies on the distance of each star to the isochrone shifted by the optimal $\mu_{Ks}$ and $E(J-Ks)$ values obtained in the previous section. Its goal is to eliminate stars that are too far from the isochrone, as they are likely field contaminants or photometric outliers that could compromise the fitting accuracy.

The procedure applies a $3\sigma$ rejection criterion along with a color tolerance radius of 0.2 mag. This threshold was chosen to be wide enough to retain genuine cluster stars but restrictive enough to remove most obvious contaminants. This strategy is particularly effective in mitigating spurious detections, as seen in Fig. \ref{f_cmd}.

Fig. \ref{f_iso-fit} shows the CMD after applying the re-decontamination process. Removed stars are shown in orange, while retained ones appear in green. The resulting distribution is cleaner and more consistent with the shifted isochrone, indicating that the process effectively reduced the contamination level.

The outcome of this stage is a dataset more representative of the GC population, significantly improving the accuracy of the final age determination.

\subsection{Age determination}
\label{ss:AGE}

With the re-decontaminated CMD and using the optimal values of $E(J-Ks)$ and $\mu_{Ks}$ together with the extinction vector $\frac{A_{Ks}}{E(J-Ks)}$ computed in Section \ref{ss_distance}, the cluster age was determined.

This process involves iterating over the original set of isochrones obtained from \texttt{PARSEC}, each with a different age between 7 and 14 Gyr, separated by intervals of 0.1 Gyr. All isochrones are shifted according to the optimal $\mu_{Ks}$ and $E(J-Ks)$ values previously determined. For each point of the shifted isochrones, bins oriented along the extinction vector are built, with a length of 0.8 mag along the vector and a width of 0.4 mag in the perpendicular direction.

For each bin, the mean position of the stars in magnitude and color is computed, and then the projected distance between this mean and the corresponding point of the shifted isochrone along the extinction vector is determined as:

\begin{align}
\label{eq:r}
R &= \frac{1}{N} \sum_{n=1}^{N} \left[ (K_s^{iso} - K_{s,n}^{obs}) \left(\frac{A_{\mathrm{Ks}}}{E(J-K_s)}\right)_{\!\!\mathrm{Ks}} \right. \nonumber \\
& \quad + \left. \left((J-K_s)^{iso} - (J-K_s)_{n}^{obs}\right) \left(\frac{A_{\mathrm{Ks}}}{E(J-K_s)}\right)_{\!\!(J-K_s)} \right] 
\end{align}

where $Ks^{iso}$ and $(J-Ks)^{iso}$ are the theoretical magnitudes and colors of each point of the shifted isochrone, $Ks_{n}^{obs}$ and $(J-Ks)_{n}^{obs}$ are the observed magnitudes and colors of the stars in the bin, $N$ is the number of stars per bin, and $\left(\frac{A_{\mathrm{Ks}}}{E(J-K\!s)}\right)_{\!\!\mathrm{Ks}}$ and $\left(\frac{A_{\mathrm{Ks}}}{E(J-K\!s)}\right)_{\!\!(J-K\!s)}$ are the components of the extinction vector along the magnitude and color directions, respectively.

The residual $R$ quantifies the relative position of the stars with respect to the shifted isochrone. If $R > 0$, the stars lie below the isochrone along the extinction vector; if $R < 0$, they lie above, in the opposite direction. The right panel of Fig. \ref{f_iso-fit2} exemplifies the calculation of $R$ for isochrones from 7 to 14 Gyr (in 1 Gyr steps) during the convergence iteration. Older ages yield positive $R$ values, indicating that stars lie mostly below the isochrone; conversely, younger ages result in predominantly negative $R$ values. A color cut ($J-Ks > 1.8$ mag) was also applied to exclude the RGB region, keeping only the subgiant sequence, which is the CMD region most sensitive to age that we were able to sample.

The behavior of $R$ for each shifted isochrone quantifies the quality of the fit between model and observations. Values of $R$ close to zero indicate good alignment between the stars and the isochrone, while larger deviations indicate a poorer fit. Therefore, the absolute mean value residual $R$, $|\overline{R}|$, is used as the fitting metric. The optimal cluster age corresponds to the minimum of $|\overline{R}|$, representing the model that best reproduces the observed stellar distribution in the CMD. The left panel of Fig. \ref{f_iso-fit2} illustrates the variation of $|\overline{R}|$ as a function of age during the convergence iteration, showing a clear minimum around 12.1 Gyr.

Once the age minimizing $|\overline{R}|$ is identified, the entire procedure is repeated starting the iteration with the corresponding isochrone to recompute new optimal values of $E(J-Ks)$ and $\mu_{Ks}$. This iterative process continues until the parameters converge to stable values, i.e., when they no longer change significantly between consecutive iterations.

To test the robustness of the method, the fitting was repeated starting from both extremes of the age range (7 and 14 Gyr). In both cases, the algorithm converged to the same final values of age, $E(J-Ks)$, and $\mu_{Ks}$,  confirming the consistency and reliability of the approach.

The final values of the parameters derived from this method are reported in Section \ref{ssec:age} of the main article, where the error analysis and astrophysical implications are discussed in detail.

\section{Cluster's orbital properties: additional information}
\label{sec:app:orb}

\begin{figure}[b!]
\centering
\includegraphics[width=\columnwidth]{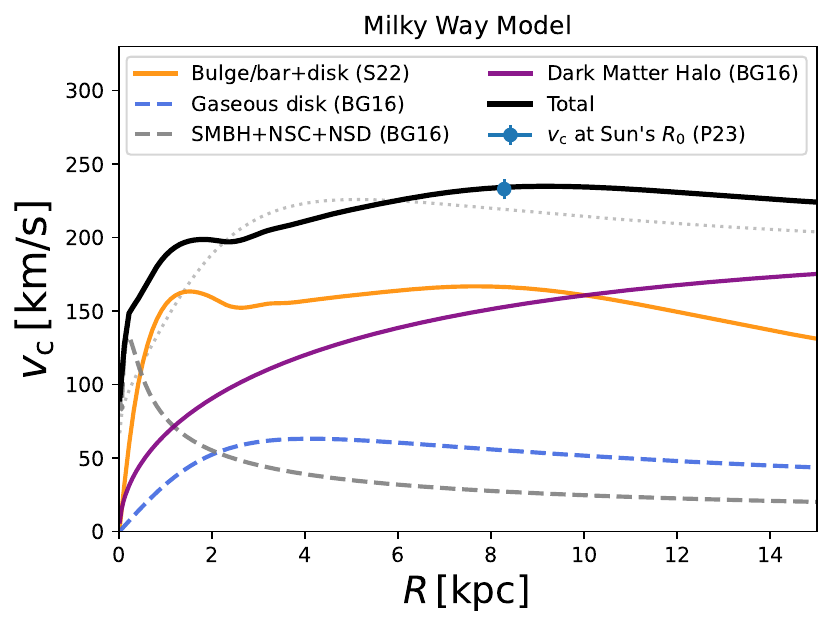}
\vspace{-0.5cm}
\caption{
Circular velocity profiles (${\rm v_{\rm c}}$) for the MW potential model used here (black curve), and its components.
We show ${\rm v_{\rm c}}$ for the stellar bar and inner disk model of \citetalias{Sormani2022} (orange).
The additional components are derived from \citetalias{Bland-Hawthorn2016} which are: the gaseous disk (blue), the nuclear components (gray) corresponding to the Sagittarius A* super massive black hole (SMBH), the nuclear star cluster (NSC), and the nuclear stellar disk (NSD), and the dark matter halo. 
More details in \S \ref{sec:app:orb}.
We include the ${\rm v_{\rm c}}$ at Sun's position ${\rm v_{\rm c}}\!=\!233\!\pm\!7\,{\rm km/s}$ \citep{Poder2023} at its galactocentric distance $R_{0}\!=\!8.2\!\pm\!0.1\, {\rm kpc}$ \citetalias{Bland-Hawthorn2016}.
We include the axisymmetric MW potential \textit{MWPotential2014} from \citet{Bovy2015} (grey dotted line) to compare the gravitational potential differences with our barred MW model.
}
\label{fig:MWvc}
\end{figure}
We updated the orbital calculator software \textsc{delorean} \citep{Blana2020} with new gravitational potentials for the Milk Way (Bla\~na et al. in prep.). 
The circular velocity profiles of the potential are shown in Fig.\ref{fig:MWvc}. 
For the MW's box/peanut bulge, flat bar, and inner disk potential we used the stellar mass density analytical model of \citetalias{Sormani2022} which was derived from made-to-measure orbital models fitted to MW 3D observations \citep{Portail2017a,Wegg2013}. 

As this model traces well the stellar mass distribution within 15 kpc,
we used it to obtain the gravitational potential by solving the Poisson equation with the Fast Fourier Transform method.
We also adopted the bar pattern speed of $\Omega_{\rm bar}\!=\!39.0\!\pm 3.5\,{\rm km\,s^{-1}\,kpc^{-1}\, rad}$ \citep{Portail2017a}, and a bar angle of $27\deg$\citep{Wegg2013} which encompasses other estimates in the literature within the errors.
We also included the potentials of other mass components with mass and scale length estimates based on \citetalias{Bland-Hawthorn2016}.
These are the nuclear components, which include the Sagittarius A$*$ supermassive black hole modeled with a Plummer profile (SMBH: $M_{\medbullet}\!=\!4.3\!\times\!10^6\,{\rm M_{\odot}}$ and a softening of $r_{\rm pl}=1.6\,{\rm pc}$), 
the nuclear star cluster (NSC: $M_{\rm NSC}\!=\!1.8\!\times\!10^7\,{\rm M_{\odot}}$ and half-mass radius $r_{\rm pl}=4.2{\rm pc}$), 
and the nuclear stellar disk modeled with a Miyamoto-Nagai profile (NSD: $M_{\rm NSD}\!=\!1.4\!\times\!10^9\,{\rm M_{\odot}}$ and scale length and height of $R_{\rm d}=90\,{\rm pc}$ and $z_{\rm h}\!=\!45\,{\rm pc}$, respectively). 
The cold gaseous disk (atomic and molecular gas) is also modeled with a Miyamoto-Nagai profile with a mass of $M_{\rm gas}\!=\!7\!\times\!10^9\,{\rm M_{\odot}}$ and scale length and height of $R_{\rm d}=2.6\,{\rm kpc}$ and $z_{\rm h}\!=\!0.3\,{\rm kpc}$, respectively.
For the dark matter halo we also use the Navarro-Frenk-White (NFW) profile adapted from \citetalias{Bland-Hawthorn2016} with a virial mass of $M_{\rm vir}\!=\!1.3\!\times\!10^{12}\,{\rm M_{\odot}}$ and concentration of $c=16$ which fits well the gas rotation curve and the circular velocity at the Sun's location (see Fig.\ref{fig:MWvc}).
We note that the hot gas and stellar halo mass contributions are smaller than the dark-halo error estimates, and therefore omitted.
We used Fourier analysis to determine the typical orbital periods of the cluster, finding values around $T_{\rm \phi}\sim100\,{\rm Myr}$, allowing us to adopt a time step of $dt\!=\!0.1\,{\rm Myr}$ to ensure enough precision while allowing faster computations.

\end{appendix}

\end{document}